\documentclass[twocolumn,showpacs,preprintnumbers,amsmath,amssymb,floatfix]{revtex4}

\usepackage{amsmath}
\usepackage{verbatim}
\usepackage{graphicx}
\usepackage{dcolumn}
\usepackage{bm}
\usepackage{url}
\usepackage{hyperref}

\usepackage{color}

\newcommand{\be}{\begin{equation}}
\newcommand{\ee}{\end{equation}}
\newcommand{\bea}{\begin{eqnarray}}
\newcommand{\eea}{\end{eqnarray}}
\newcommand{\nn}{\nonumber}
\newcommand{\dee}{\mathrm{d}}
\newcommand{\Mc}{{\cal M}}
\newcommand{\Ms}{M_{\odot}}

\begin{document}

\title{Constraining the neutron star equation of state with gravitational wave signals from coalescing binary neutron stars}

\author{M.~Agathos$^{1}$}
\email{magathos@nikhef.nl}
\author{J.~Meidam$^{1}$}
\author{W.~Del Pozzo$^{2}$}
\author{T.G.F.~Li$^{3}$}
\author{M.~Tompitak$^{1,4}$}
\author{J.~Veitch$^{2}$}
\author{S.~Vitale$^{5}$}
\author{C.~Van Den Broeck$^{1}$}
\affiliation{$^1$Nikhef -- National Institute for Subatomic Physics, Science Park 105, 1098 XG Amsterdam, The Netherlands \\
$^2$School of Physics and Astronomy, University of Birmingham, Edgbaston, Birmingham B15 2TT, United Kingdom \\
$^3$LIGO Laboratory, California Institute of Technology, Pasadena, CA 91125, USA \\
$^4$Leiden Institute of Physics, Leiden University, Niels Bohrweg 2, 2333 CA Leiden, The Netherlands \\
$^5$LIGO Laboratory, Massachusetts Institute of Technology, Cambridge, MA 02139, USA}

\date{\today}

\begin{abstract}
Recently exploratory studies were performed on the possibility of constraining the neutron star equation of state (EOS) using signals from coalescing binary neutron stars, or neutron star-black hole systems, as they will be seen in upcoming advanced gravitational wave detectors such as Advanced LIGO and Advanced Virgo. In particular, it was estimated to what extent the combined information from multiple detections would enable one to distinguish between different equations of state through hypothesis ranking or parameter estimation. Under the assumption of zero neutron star spins both in signals and in template waveforms and considering tidal effects to 1 post-Newtonian (1PN) order, it was found that $\mathcal{O}(20)$ sources would suffice to distinguish between a stiff, moderate, and soft equation of state. Here we revisit these results, this time including neutron star tidal effects to the highest order currently known, termination of gravitational waveforms at the contact frequency,  neutron star spins, and the resulting quadrupole-monopole interaction. We also take the masses of neutron stars in simulated sources to be distributed according to a relatively strongly peaked Gaussian, as hinted at by observations, but without assuming that the data analyst will necessarily have accurate knowledge of this distribution for use as a mass prior. We find that especially the effect of the latter is dramatic, necessitating many more detections to distinguish between different EOS and causing systematic biases in parameter estimation, on top of biases due to imperfect understanding of the signal model pointed out in earlier work. This would get mitigated if reliable prior information about the mass distribution could be folded into the analyses.   
\end{abstract}


\maketitle

\section{Introduction}

Second-generation ground-based interferometric gravitational wave (GW) detectors are currently under construction: Advanced LIGO  \cite{Harry:2010zz} in the US, Advanced Virgo \cite{TheVirgo:2014hva} in Italy, and KAGRA \cite{Somiya:2011np} in Japan. GEO-HF in Germany \cite{Luck:2010rt,Affeldt:2014rza} is already taking data. Later in the decade, LIGO-India \cite{Iyer:2011LIGOIndia} may join this global network of observatories. Among the most promising sources for a first direct detection of gravitational waves are compact binaries composed of neutron stars or black holes, with detection rates in the range $1 - 100\,\mbox{yr}^{-1}$ depending on the astrophysical event rate, the instruments' duty cycle, and the sensitivities of the detectors \cite{Abadie:2010cf,Aasi:2013wya}; see also \cite{Chen:2012qh} for detection rates under the assumption that short, hard gamma ray bursts are caused by coalescing binaries.  

Coalescing binaries consisting of two neutron stars (BNS), a neutron star and a black hole (NSBH), or two black holes (BBH) have rich scientific potential. They can be used to test general relativity in the genuinely strong-field regime \cite{Li:2011cg,Li:2011vx,Agathos:2013upa} and are self-calibrating ``standard sirens'' for cosmology \cite{Schutz:1986gp,Nissanke:2009kt,DelPozzo:2012zz}. Moreover, BNS and NSBH coalescences can be used as probes of the elusive neutron star equation of state (EOS), about which little is currently known \cite{Lattimer:2012nd}. 

The possibility of constraining or measuring the neutron star EOS with gravitational wave observations of BNS coalescences has recently been the subject of extensive investigation. The way in which the EOS enters the GW signal from a coalescing binary is mainly through tidal deformation. During the last stages of inspiral, the tidal field $\mathcal{E}_{ij}$ of one component of a binary will induce a quadrupole moment $Q_{ij}$ in the other, where to leading order in the adiabatic approximation $Q_{ij} = -\lambda(\mbox{EOS}; m)\,\mathcal{E}_{ij}$.  The tidal deformability $\lambda(\mbox{EOS}; m)$ depends on the neutron star mass $m$ in a way that is governed by the EOS. This  deformation of the neutron stars has an effect on the orbital motion, and hence on the waveform of the emitted gravitational wave signal; in particular, it enters the phase $\Phi(t)$. The deformability is related to the radius $R(m)$ through  $\lambda(m) = (2/3) k_2(m) R^5(m)$, where $k_2$ is the second Love number. 
Although tidal effects enter the phase at high apparent post-Newtonian order (first appearing alongside the 5 post-Newtonian (5PN) phase contribution), these corrections come with a large prefactor: $\lambda(m)/M^5 \propto (R/M)^5 \sim 10^2 - 10^5$ (with $M$ the total mass of the binary), so that they may be observable with advanced detectors. We stress that there are other ways in which the EOS enters the gravitational waveform; we will come to these momentarily.

Read \emph{et al.}~estimated that with a single close-by source (at a distance of $\sim 100$ Mpc), the neutron star radius could be constrained to 10\% \cite{Read:2009yp}. Hinderer \emph{et al.}~performed a Fisher matrix analysis with PN waveforms truncated at 450 Hz to see how well the neutron star tidal deformability might be measurable from the low-frequency inspiral dynamics alone \cite{Hinderer:2009ca}, concluding that it would be difficult to extract much information from this regime, at least with second-generation detectors. Damour, Nagar, and Villain performed a Fisher matrix calculation using effective one-body waveforms up to the point where the neutron stars are touching each other \cite{Damour:2012yf}, which suggested that it might be possible after all to gain information about the EOS with advanced detectors. Lackey \emph{et al.}~performed a similar analysis for NSBH, which also indicated encouraging prospects \cite{Lackey:2011vz}. The abovementioned work considered single detections; a first study for multiple detected sources was performed by Markakis \emph{et al.}~\cite{Markakis:2010mp}, who concluded that a similar accuracy to \cite{Read:2009yp} could be attained with 3 sources that have low signal-to-noise ratio (SNR). On the other hand, Fisher matrix estimates can be unreliable at low SNR \cite{Cokelaer:2008zz,Vallisneri:2007ev,Zanolin:2009mk}, prompting more in-depth assessments. 

The first fully Bayesian investigation of the problem, in a realistic data analysis setting, was performed by Del Pozzo \emph{et al.}~\cite{DelPozzo:2013ala}. BNS signals were ``injected'' into simulated detector noise, assuming the projected final design sensitivity of the Advanced LIGO-Virgo detector network. Sources were distributed in an astrophysically realistic way, leading to the distribution of SNRs that we expect to see towards 2018. Two different Bayesian analysis methods were employed: hypothesis ranking within a list of different theoretical EOS, and parameter estimation. The former trivially allows one to combine information from multiple sources so as to arrive at a stronger result. To do the same with the latter method, parameters need to be identified that do not vary from source to source; in \cite{DelPozzo:2013ala} these were simply taken to be coefficients in a Taylor expansion of $\lambda(m)$ in powers of $(m-m_0)/M_\odot$, where $m_0$ is some reference mass. A similar analysis to \cite{DelPozzo:2013ala} in terms of parameter estimation was recently performed by Lackey and Wade \cite{Lackey:2014fwa}. The latter authors modeled the EOS as piecewise polytropes, allowing them to directly arrive at statements on the measurability of pressure as a function of density and neutron star radius as a function of mass. The latter method also has the advantage that physical priors such as causality can more easily be folded in. 
Both \cite{DelPozzo:2013ala} and \cite{Lackey:2014fwa} concluded that $\lambda(m_0)$, with $m_0 = 1.4\, M_\odot$, could be measured with an accuracy of $\sim 10\%$ by combining information from a few tens of sources.

Of necessity, the studies in \cite{DelPozzo:2013ala,Lackey:2014fwa} used relatively simple waveform approximants, as otherwise the simulated data analysis problem would have been intractable with existing methods and computational infrastructure. Much effort is being put into large-scale numerical simulations of the spacetimes of coalescing BNS, especially of the late inspiral \cite{Baiotti:2010xh,Baiotti:2011am,Bernuzzi:2012ci,Hotokezaka:2013mm,Read:2013zra,Radice:2015nva}. The resulting waveforms are ``hybridized'' by matching them onto post-Newtonian or effective one-body waveforms, so that the earlier inspiral is also represented. While such waveforms represent the state of the art in our understanding of BNS coalescence, producing a single one of them can take weeks. By contrast, high quality parameter estimation requires millions of waveforms to be compared with the data (see \cite{Veitch:2014wba} and references therein). A full solution of the problem of inferring the EOS from BNS detections will likely involve a combination of constructing phenomenological or ``tuned'' waveform models with input from numerical relativity  \cite{Sturani:2010yv,Sturani:2010ju,Pan:2013rra,Taracchini:2013rva,Schmidt:2012rh,Hannam:2013oca,Schmidt:2014iyl}, and significantly speeding up the analysis of the data, \emph{e.g.}~through the use of Reduced Order Modeling; see \cite{Purrer:2014fza,Canizares:2014fya} and references therein. In that regard we note the recent work by Bernuzzi \emph{et al.}~\cite{Bernuzzi:2014owa}, who derived an effective one-body model that accurately describes tidal effects close to merger for a number of different EOSs, matching results from numerical simulations essentially to within the numerical uncertainties.

On the other hand, when focusing on the inspiral regime, since the way tidal effects contribute to the phase is analogous for all of the PN approximants and the effective one-body waveforms \cite{Vines:2011ud,Vines:2010ca,Damour:2012yf}, one might think that existing waveforms would already suffice to reliably extract information on the EOS from this part of the signal. However, as pointed out in \cite{Hinderer:2009ca} and studied in detail in \cite{Favata:2013rwa,Yagi:2013baa,Wade:2014vqa,Lackey:2014fwa}, significant biases can arise in the estimation of EOS effects due to discrepancies between waveform approximants -- and presumably between these approximants and the true signal waveform -- at high frequencies. Much of this is due to the fact that for the underlying \emph{point particle} waveforms, the different PN waveforms and the effective one-body models differ significantly from each other at frequencies $f \gtrsim 400$ Hz, where tidal effects become apparent. 

An important observation was made by Read \emph{et al.}~ \cite{Read:2013zra}, who studied the ``distinguishability'' $|| \delta h || \equiv \sqrt{\langle h_2 - h_1 | h_2 - h_1 \rangle}$ in terms of the usual PSD-weighted inner product $\langle \, \cdot \, | \, \cdot \,\rangle$ for waveforms $h_1$, $h_2$ of the same family but differing in their parameter values, in this case $\lambda$. As can be seen in their Fig.~12, the dependence of $||\delta h||$ on changes in $\lambda$ is very similar for PN approximants and for hybridized numerical waveforms. Thus one may anticipate that PN approximants will allow us to \emph{predict} how well one will be able to infer the EOS from GW measurements when sufficiently accurate waveforms will eventually become available for use in data analysis algorithms, even if the latter waveforms and appropriate analysis techniques are not yet at our disposal today. This will then inform the waveform modeling and data analysis communities as to what can reasonably be expected in terms of scientific output once their considerable efforts have come to fruition. Providing such an assessment is the aim of the present paper.

In this work we significantly expand on the results of \cite{DelPozzo:2013ala,Lackey:2014fwa} in two ways: (a) we provide much more extensive statistics through analysis of a larger number of simulated BNS sources, so as to identify worst-case and best-case scenarios depending on the detector noise realizations and the astrophysical distribution of the detected sources; and (b) we take into account as many physical effects as have been modeled, including tidal effects to highest known order \cite{Damour:2012yf}, neutron star spins, the quadrupole-monopole interaction \cite{Poisson:1997ha,Laarakkers:1997hb}, the impact of early waveform termination due to the finite neutron star radii, and a relatively strongly peaked Gaussian distribution for the neutron star masses that is expected on astrophysical grounds \cite{Valentim:2011vs,Kiziltan:2010ct,Kiziltan:2011mj,Kiziltan:2013oja}. The effect of the latter was ignored in previous work, but as we shall see, it has a significant impact on inference of the EOS. 
As in \cite{DelPozzo:2013ala}, we use two Bayesian data analysis methods: hypothesis ranking within a set of possible EOSs to see which one the true EOS is closest to, and parameter estimation on coefficients in a Taylor expansion of the tidal deformability.
We find that when source component masses are in a narrow distribution, using a flat component mass prior for the analyses (as in previous work) increases the number of detections needed to distinguish between a soft, moderate, and stiff equation of state, and causes additional biases in parameter estimation on top of the ones due to imperfect knowledge of the signal model. On the other hand, this would get mitigated if we could assume accurate knowledge of the astrophysical mass distribution for neutron stars in binaries, so that it could be used as the prior distribution of masses.

This rest of this article is structured as follows. In Sec.~\ref{sec:waveforms} we introduce the waveform model and the EOS-related contributions from the effects mentioned above. In Sec.~\ref{sec:methods} we explain the two main methods used in the simulated data analysis: hypothesis ranking and parameter estimation. Sec.~\ref{sec:setup} explains the setup of our simulations. In Sec.~\ref{sec:results} we show the main results of this paper. A summary and discussion is given in Sec.~\ref{sec:discussion}. Finally, in the Appendix we further investigate the impact of the prior on component masses.

Throughout this paper we will use units such that $G = c = 1$ unless stated otherwise.

\section{Waveform model and effects of the neutron star equation of state}
\label{sec:waveforms}

In this section we first discuss the general form of our waveform model, and then the way in which EOS effects enter. 

\subsection{General form of the waveform model}

We model gravitational waveforms from the quasicircular inspiral of BNS systems using the stationary phase approximation (SPA), which yields a convenient analytic expression of the observed GW strain in the frequency domain \cite{1987thyg.book..330T,Sathyaprakash:1991mt}:
\be
\tilde{h}(f) = \frac{1}{D}\frac{\mathcal{A}(\theta, \phi, \iota, \psi, \Mc, \eta)}{\sqrt{\dot{F}(f; \Mc, \eta, \chi_1, \chi_2)}}\,f^{2/3}\,e^{i\Psi(f; t_c, \varphi_c, \Mc, \eta, \chi_1, \chi_2)}.
\label{hoff}
\ee
Here $D$ is the distance to the source; $(\theta, \phi)$ denote the sky position with respect to the interferometer; $(\iota, \psi)$ determine the orientation of the orbital plane in relation to the observer; $\Mc = M \eta^{3/5}$ is the \emph{chirp mass}, with $M = m_1 + m_2$ being the total mass and $\eta = m_1 m_2/M^2$ the symmetric mass ratio; $t_c$ and $\varphi_c$ are, respectively, the time and the phase at coalescence; and $\chi_1$, $\chi_2$ are the neutron stars' dimensionless spins.  The ``frequency sweep'' $\dot{F}(f; \Mc, \eta, \chi_1, \chi_2)$ (related to the time domain phase $\Phi(t)$ by $\dot{F}(f) = \ddot{\Phi}(t(f))/\pi$) is an expansion in powers of $f$ with coefficients that depend on masses and spins, and the SPA phase takes the general form
\be
\Psi(f) = 2\pi f t_c - \varphi_c - \frac{\pi}{4} + \sum_k \left[ \psi_k + \psi^{(l)}_k \ln f \right] f^{(k-5)/3},
\label{Psioff}
\ee
where the $\psi_k$ and $\psi_k^{(l)}$ again depend on masses and spins. For the low-mass systems considered in this paper, the Advanced LIGO-Virgo network will not be very sensitive to subdominant PN contributions to the amplitude \cite{VanDenBroeck:2006qu,VanDenBroeck:2006ar,O'Shaughnessy:2013vma}, and we use the ``restricted'' post-Newtonian approximation, in which only leading-order PN contributions to the amplitude are taken into account. In particular, this means that spin and EOS effects appear in the phasing only. 

For the purposes of this paper, the phase was taken to 3.5PN order, with inclusion of spin effects up to 2.5PN following \cite{Arun:2008kb}, and we refer to that paper for explicit expressions. For simplicity we assume that the components'  spins are aligned or anti-aligned with each other and with the direction of orbital angular momentum; at the time this work was started, frequency-domain, precessing-spin waveform approximants like the ones of Lundgren and O'Shaughnessy \cite{Lundgren:2013jla}, of Hannam \emph{et al.}~\cite{Hannam:2013oca,Schmidt:2014iyl}, and of Klein \emph{et al.}~\cite{Klein:2014bua} were not yet available. 
It is quite possible that inclusion of precession would aid in breaking degeneracies between spins and mass ratio, as was suggested in \emph{e.g.}~\cite{Chatziioannou:2014coa}, enabling more accurate measurements of EOS effects. 

We take into account three ways in which the EOS affects the waveform: tidal deformations, the quadrupole-monopole effect, and the possible early termination of the waveform due to the finite size of the neutron stars, whose radii are set by their masses and the EOS. Let us discuss these in turn.

\subsection{Tidal deformations}
\label{sec:tidal}

Towards the end of the evolution of a BNS system, when the gravitational wave frequency reaches $f \gtrsim 400$ Hz \cite{Hinderer:2009ca}, the tidal tensors $\mathcal{E}_{ij}$ of one component of the binary will start to induce a significant quadrupole moment $Q_{ij}$ in the other. In the adiabatic approximation, the two are related by \cite{Thorne:1997kt,Flanagan:2007ix,Vines:2011ud}
\be
Q_{ij} = -\lambda(m) \, \mathcal{E}_{ij},
\ee 
where $m$ is the mass of the neutron star that is experiencing the quadrupole deformation, and the function $\lambda(m)$ is the tidal deformability, which is determined by the EOS. The deformations of the two neutron stars in turn affect the orbital motion, and this is one way in which the EOS gets imprinted upon the gravitational waveform. The deformability $\lambda(m)$ is related to the second Love number $k_2(m)$ and the neutron star radius $R(m)$ through $\lambda(m) = (2/3)\,k_2(m)\,R^5(m)$. Tidal effects only enter the phase starting at 5PN order \cite{Flanagan:2007ix}, but as mentioned before, the prefactors are sizable ($\lambda/M^5 \propto (R/M)^5 \sim 10^2 - 10^5$), which is why we can hope to infer information on the EOS from the tidal deformation. 

The effects of tidal deformations on the orbital motion were calculated up to 1PN (or 6PN in phase) by Vines, Flanagan, and Hinderer \cite{Vines:2011ud}, and more recently to 2.5PN (or 7.5PN in phase) by Damour, Nagar, and Villain \cite{Damour:2012yf}. The latter expression is what we will be using in this paper; for completeness we reproduce it here. In terms of the characteristic velocity $v = (\pi M f)^{1/3}$, one has
\be
\Psi(v) = \Psi_{\rm PP}(v) + \Psi_{\rm tidal}(v),
\ee
where $\Psi_{\rm PP}(v)$ is the phase for the inspiral of point particles, and $\Psi_{\rm tidal}(v)$ is the contribution from tidal effects. The latter takes the form
\begin{widetext}
\bea
\Psi_{\rm tidal}(v) &=& \frac{3}{128 \eta} v^{-5} \sum_{A=1}^{2} \frac{\lambda_A}{M^5 X_A} \left[ - 24 \left( 12 - 11X_A \right)\, v^{10} + \frac{5}{28} \left( 3179 - 919 X_A  - 2286 X_A^2 + 260 X_A^3  \right) v^{12} \right. \nonumber\\
&& \left. + 24 \pi (12 - 11 X_A) v^{13}  \right. \nonumber\\
&& \left. - 24 \left( 
\frac{39927845}{508032} - \frac{480043345}{9144576} X_A + \frac{9860575}{127008} X_A^2 - \frac{421821905}{2286144} X_A^3 + \frac{4359700}{35721} X_A^4 - \frac{10578445}{285768} X_A^5 \right) v^{14} \right.  \nonumber\\
&& \left. + \frac{\pi}{28}  \left( 27719 - 22127 X_A + 7022 X_A^2 - 10232 X_A^3 \right) v^{15}
\right],
\label{tidal}
\eea
\end{widetext}
where $X_A = m_A/M$, $A  =1, 2$, and $\lambda_A = \lambda(m_A)$. We should note that the calculation leading to this expression ignores (i) contributions from higher-order multipoles as these are estimated to give small corrections, and (ii) a number of as yet unknown functions that appear in the 7PN phase contribution; in \cite{Damour:2012yf} these too were argued to be negligible and we refer to that paper for details. 
Contributions to the phase at increasing PN order, for a BNS system of $(1.35,1.35)\Ms$ with a stiff (MS1) EOS, are illustrated in Fig.~4. 

For the function $\lambda(m)$, in our simulated signals we will use quartic polynomial fits to predictions corresponding to different EOS from Hinderer \emph{et al.}~\cite{Hinderer:2009ca}, with maximum residuals of $\sim 0.02$ (which will turn out to be negligible compared to the measurability of $\lambda$). Examples of such fits for a soft (labeled SQM3), a moderate (H4), and a stiff EOS (MS1) are shown in Fig.~1.

\begin{figure}
\label{fig:lambda}
\includegraphics[width=\columnwidth]{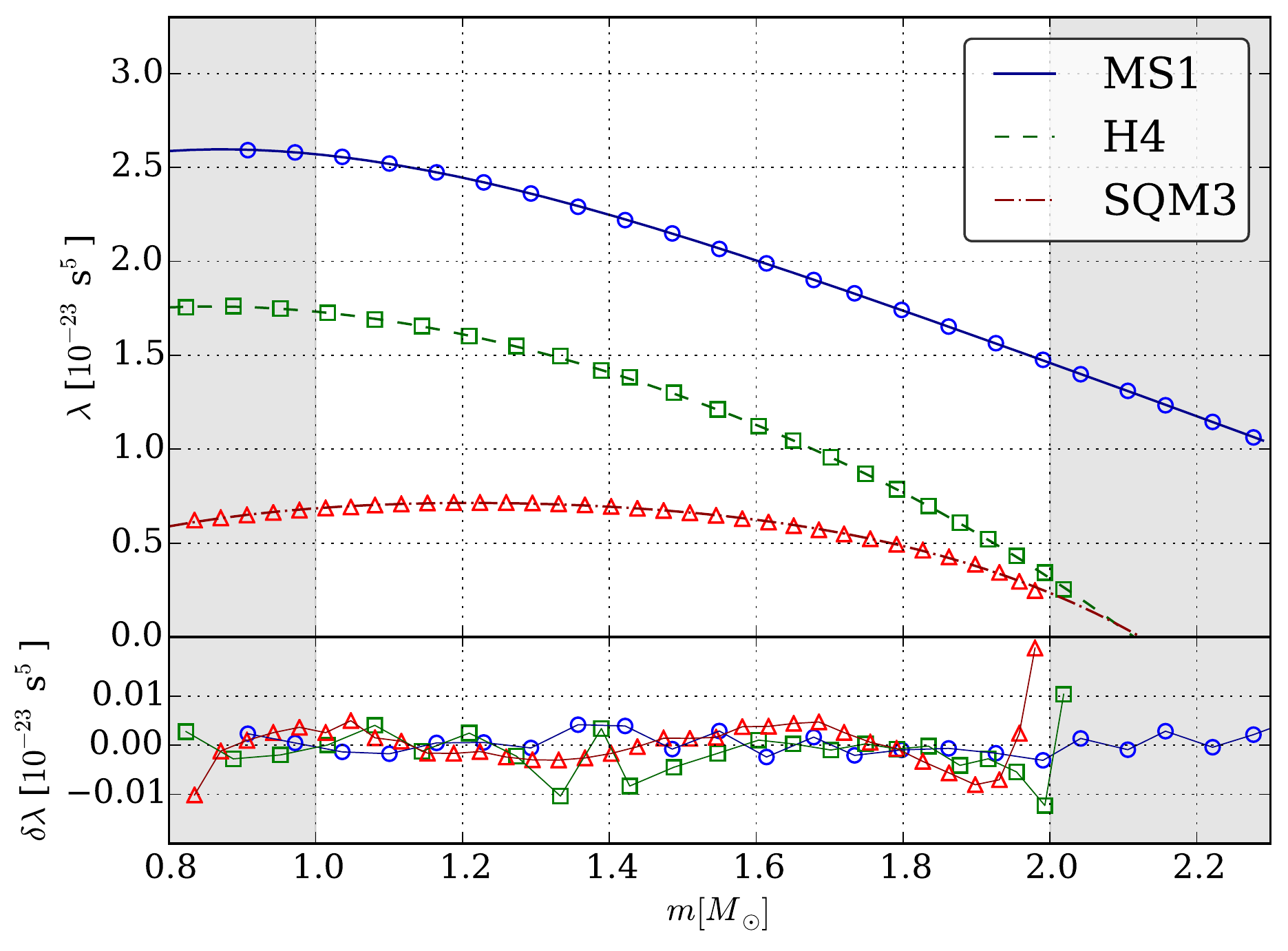}
\caption{The tidal deformability parameter $\lambda(m)$ as a function of neutron star mass for three different EOS: a soft one (SQM3), a moderate one (H4), and a stiff one (MS1). Adapted from \cite{Hinderer:2009ca}. Curves are fitted quartic polynomials, whose residuals are shown in the lower subplot. Only masses within the unshaded region $[1,2]\Ms$ will be considered in our analyses.}
\end{figure}

\subsection{Quadrupole-monopole effects}

As mentioned before, tidal effects are not the only way the EOS enters into the gravitational waveform. If a neutron star is spinning, it takes on an oblate shape. Assuming an axisymmetric mass distribution with respect to the axis of rotation, the deformation can be expressed to leading order by means of a dimensionless quadrupole moment parameter $q$, defined as~\cite{Laarakkers:1997hb}
\begin{equation}
q = -\frac{5}{2} \lim_{r\rightarrow \infty} \left(\frac{r}{M}\right)^3 \int_{-1}^{1} \nu(r,\theta) P_2(\cos\theta)\,  \dee\cos\theta , 
\end{equation}
where $P_2(x) = (3x^2-1)/2$ is the second Legendre polynomial, and $\nu$ is a potential related to the metric of a stationary axially symmetric body; more specifically, the line element in the form introduced by Komatsu-Eriguchi-Hachisu~\cite{Komatsu:1989zz} reads:
\begin{align}
\dee s^{2} = - e^{-2\nu} \dee t^{2} &+ r^{2} \sin^{2}\theta \: e^{2\beta} \left(\dee\phi - \omega \dee t \right)^{2} \nn \\
 &+ e^{2\alpha} \left( \dee r^{2} + r^{2} \dee \theta^{2} \right),
\end{align}
where the undetermined $\alpha, \beta, \nu$ are all functions of $(r,\theta)$.
The quadrupole moment $q$ is the leading-order ($1/r^{3}$) coefficient of the second multipole in the asymptotic expansion of $\nu(r,\theta)$ and can be calculated numerically.
This quantity is the general-relativistic equivalent of the Newtonian mass quadrupole moment.

Since a stiffer EOS implies a larger neutron star (NS) radius for a given mass, the quadrupole moment increases in absolute value with the stiffness of the EOS. 
Examples of $q$ estimates for different EOS were calculated numerically in~\cite{Laarakkers:1997hb} based on the expressions of Ryan~\cite{Ryan:1995wh,Ryan:1996nk}.
These demonstrated the dependence on the dimensionless spin $\chi$, which for a fixed NS mass can be fit very well up to the maximum spin value $\chi_{\rm max}$ (also dependent on the EOS) by a quadratic rule:
\begin{equation}
q \simeq - a \chi^2,
\label{quadraticrule}
\end{equation}
where $a = a_{\rm EOS}(m)$ is a mass-dependent parameter. 
Further evidence to support the quadratic relation Eq.~(\ref{quadraticrule}) is given in~\cite{Pappas:2012ns,Pappas:2012qg}.
The authors of \cite{Pappas:2012ns,friedman+stergioulas} also point out a spin correction in the identification of multipole moments that was previously overlooked; this correction preserves the quadratic spin behaviour of Eq.~(\ref{quadraticrule}), and vanishes in the slow-rotation limit.
Assuming that this relation will hold for any EOS, we will only be concerned with the spin-independent parameter $a$ which, similar to the tidal deformability parameter $\lambda$, has a functional dependence on the neutron mass that is determined by the EOS.

The effect of such a quadrupole moment on the gravitational waveform emitted by a binary system was derived in \cite{Poisson:1997ha}.
To Newtonian order it introduces an additional coupling in the effective gravitational potential, between the mass quadrupole of each spinning neutron star and the mass of its companion, whence the name ``quadrupole-monopole (QM) effect''. In the stationary phase approximation, the additional contribution to the GW phase due to the QM interaction reads:
\be
\Psi_{\rm QM}(v) =  - \frac{30}{128 \eta} \sigma_{\rm QM} v^{-1},
\label{QM}
\ee
making it of 2PN order in phase. The parameter $\sigma_{\rm QM}$ depends on masses and spins through 
\begin{align}\label{sigmaQM}
\sigma_{\rm QM} = & - \frac{5}{2} \sum_{A=1,2} q_A \left(\frac{m_A}{M}\right)^2 \left[ 3 (\hat{\chi}_A \cdot \hat{L})^2 - 1 \right] \\
\simeq & \frac{5}{2} \sum_{A=1,2} a(m_A) \left(\frac{m_A}{M}\right)^2 \left[ 3 (\hat{\chi}_A \cdot \hat{L})^2 - 1 \right]\,\chi_A^2 \, , \nn
\end{align}
where the unit vectors $\hat{\chi}_A$ are the direction of the spins. In the last line we used the rule (\ref{quadraticrule}); we see that with this assumption, $\Psi_{\rm QM}(v)$ is quadratic in the component spins. Finally, note that in the case of (anti-)aligned spins, which we will assume throughout, $3 (\hat{\chi}_A \cdot \hat{L})^2 - 1 = 2$.  

As mentioned above, in our simulations we will use predictions for $\lambda(m)$ corresponding to different EOSs from \cite{Hinderer:2009ca}. In order to compute $a(m)$, we make use of the recently discovered phenomenological \emph{Love-Q} relation \cite{Yagi:2013bca,Yagi:2013awa}, which is believed to hold irrespective of the EOS:
\begin{align}
\ln a(m) & = 0.194 + 0.0936\, \ln \frac{\lambda}{m^5} + 0.0474\, \left( \ln \frac{\lambda}{m^5} \right)^2 \nn\\
-& 4.21\times 10^{-3} \left( \ln \frac{\lambda}{m^5} \right)^3 + 1.23 \times 10^{-4} \left( \ln \frac{\lambda}{m^5} \right)^4. \nonumber\\
\label{aofm}
\end{align}
The relative fractional errors due to the universal fit were estimated in~\cite{Yagi:2013awa} for several EOSs to be at the $1\%$ level.
Together with Eqs.~(\ref{sigmaQM}) and (\ref{QM}), this then allows us to compute the QM contribution to the phase. Fig.~2 shows $a(m)$ for the EOSs in Fig.~1.
QM contributions to the phase are expected to be subdominant compared to the tidal effects of Sec.~\ref{sec:tidal}, even for relatively fast spinning NS, as shown in Fig.~4. 

\begin{figure}
\label{fig:a}
\includegraphics[width=\columnwidth]{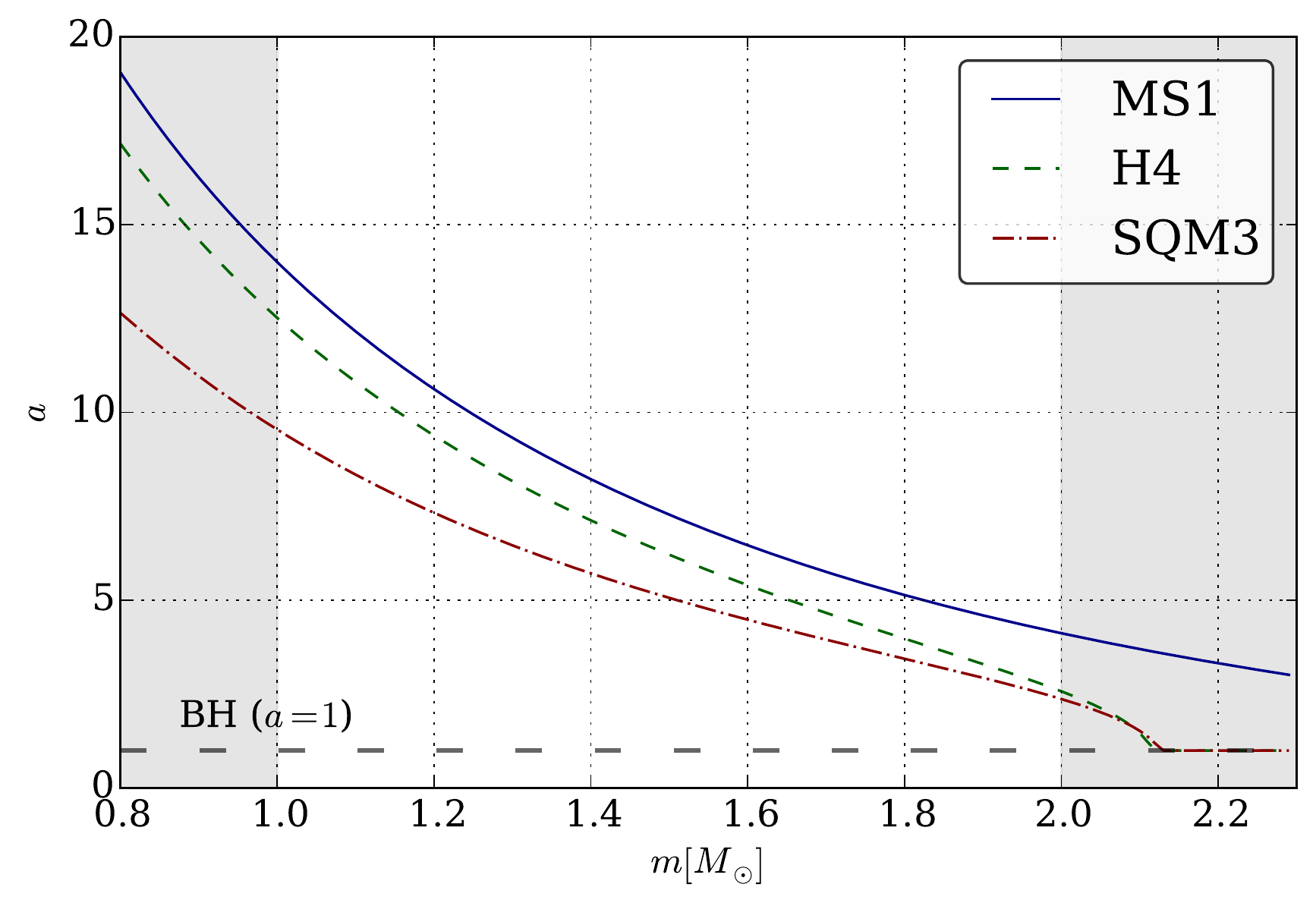}
\caption{The quadrupole parameter $a(m)$ as a function of neutron star mass for the three different EOSs in Fig.~1. 
The horizontal dashed line indicates the value for black holes, which is $a = 1$~\cite{RevModPhys.52.299}.
Only masses within the unshaded region $[1,2]\Ms$ will be considered in our analyses.}
\end{figure}

\subsection{Termination of the waveform at contact}

In the recent simulations \cite{DelPozzo:2013ala,Lackey:2014fwa}, the waveform was cut off at a frequency corresponding to the last stable circular orbit (LSO) in the point particle limit, given by
\be
f_{\rm LSO} = \frac{1}{6^{3/2} \pi M}.
\ee
However, as we shall see below, it will often happen that the two neutron stars attain physical contact before the corresponding distance between the components is reached. In this paper, we instead impose the cutoff 
\be
f_{\rm cut} = \mbox{min} \{ f_{\rm LSO}, f_{\rm contact} \},
\label{termination}
\ee
where, using Kepler's third law, the ``contact frequency'' is given by
\be
f_{\rm contact} = \frac{1}{\pi} \left( \frac{M}{R(m_1) + R(m_2)} \right)^{1/2}.
\label{fcontact}
\ee

We stress that the termination condition (\ref{termination}) is still a heuristic one, but it will be more realistic than termination at $f_{\rm LSO}$. Moreover, the length of the waveform itself carries physical information \cite{Mandel:2014tca}, in this case on the EOS, which we wish to incorporate~\footnote{It seems reasonable to expect that termination at our $f_{\rm cut}$ will be sufficiently indicative of the in reality more complicated but nevertheless dramatic changes in the waveform evolution that will occur around that frequency, and which should indeed carry information about the EOS \cite{Baiotti:2010xh,Baiotti:2011am,Bernuzzi:2012ci,Hotokezaka:2013mm,Read:2013zra,Radice:2015nva}.}. On the other hand, shorter waveforms have a smaller number of cycles from which information can be extracted; when we come to the results of our simulations we will see which effect wins out.

\begin{figure}
\label{fig:fend}
\includegraphics[width=\columnwidth]{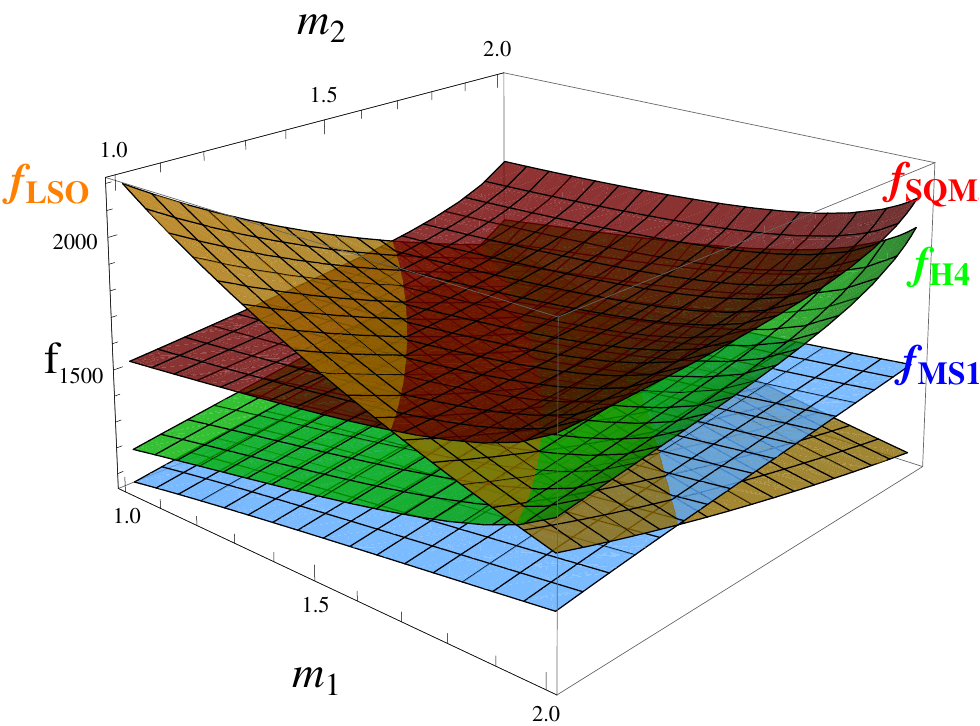} 
\caption{The frequencies $f_{\rm LSO}$ and $f_{\rm contact}$ as functions of $m_1$, $m_2$ for the EOS shown in Fig.~1.}
\end{figure}

In order to compute the radii $R(m_1)$, $R(m_2)$, we again make use of a recently discovered phenomenological relation, this time between the compactness $\mathcal{C} = m/R$ and $\lambda$ \cite{Maselli:2013mva}:
\be
\mathcal{C} = 0.371 - 3.91 \times 10^{-2} \ln \frac{\lambda}{m^5} + 1.056 \times 10^{-3} \left( \ln \frac{\lambda}{m^5} \right) ^2.
\label{compactness}
\ee
For a given EOS (\emph{i.e}~a given relationship $\lambda(m)$), the above expression gives us $R(m)$, from which the contact frequency (\ref{fcontact}) is obtained.
The relative error in the compactness (and hence in the radius) due to the fit of Eq.~(\ref{compactness}) was found to be at the $2\%$ level, implying a similar error in the contact frequency.

Fig.~3 shows the dependence of $f_{\rm LSO}$ and $f_{\rm contact}$ on component masses $m_1$, $m_2$ for the EOS considered above. Note how in the astrophysically relevant range $m_A \in [1,2]\,M_\odot$, $A = 1, 2$, it often happens that $f_{\rm contact} < f_{\rm LSO}$, especially for low masses and for the stiffer EOS (MS1) which can support larger neutron star radii.


\begin{figure}
\label{fig:PhaseTidalQM}
\includegraphics[width=\columnwidth]{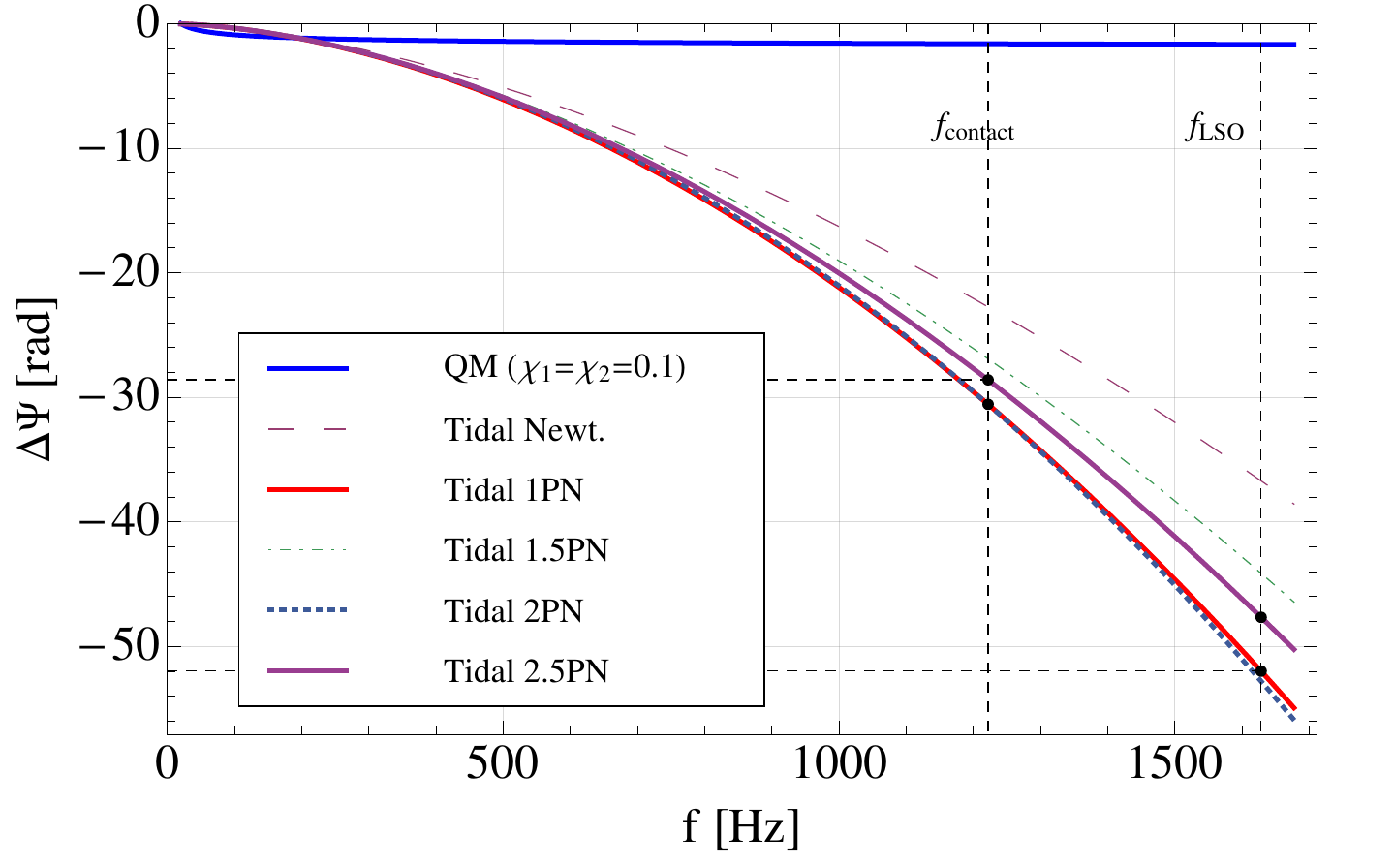} 
\caption{Phase contributions of the QM effect and tidal effects up to different PN orders as functions of GW frequency for a $(1.35,1.35)\Ms$ binary with a stiff EOS (MS1). The QM contribution from each NS scales quadratically with its spin and is shown here for $\chi_1=\chi_2=0.1$. The dashed vertical lines indicate the contact and LSO frequencies. }
\end{figure}

\section{Bayesian methods for inferring the neutron star equation of state}
\label{sec:methods}

In this section we present two qualitatively different Bayesian methods that one may use to acquire information on the neutron star equation of state: (i) hypothesis ranking for  different proposed EOS based on how well each of them matches the available data, and (ii) the estimation of parameters which for a given EOS will be the same across sources. Both of these allow us to combine information from multiple detections so as to arrive at a stronger result. These methods were already explained in \cite{DelPozzo:2013ala}; for completeness we recall the basic ideas.

\subsection{Hypothesis ranking}
\label{sec:MS}

Given a set of (finitely many) EOS models $\{M_1, M_2, \ldots, M_K\}$, we will be interested in ranking them in the light of the available data. The ranking process will be on a set of hypotheses $\{H_i; i=1, \ldots, K \}$, where $H_i$ states that $M_i$ is the true model for the neutron star EOS. Each model comes with a particular functional dependence of the tidal deformability on the mass, $\lambda^{(i)}(m)$, and hence a QM parameter $a^{(i)}(m)$ and radius-mass relation $R^{(i)}(m)$ obtained through Eqs.~(\ref{aofm}) and (\ref{compactness}), respectively. The general form of the waveform given by Eqs.~(\ref{hoff}) and (\ref{Psioff}), together with the tidal and QM contributions (\ref{tidal}), (\ref{QM}) and the waveform termination condition (\ref{termination}), yields a waveform model $\tilde{h}^{(i)}(f; \vec{\theta})$ associated with the hypothesis $H_i$. The parameter spaces $\{\vec{\theta}\}$ (masses, spins, sky position, orientation, distance, time at coalescence, and phase at coalescence) are the same for all the hypotheses $H_i$, $i = 1, \ldots, K$, but for given component masses $m_A$, $A = 1, 2$, the calculated values for $\lambda_A = \lambda^{(i)}(m_A)$, $a_A = a^{(i)}(m_A)$, and $f^{(i)}_{\rm cut}$ differ between one hypothesis and the next. 

Given a detection $d$ (to be thought of as a stretch of detector data containing a confirmed BNS signal), a hypothesis $H_i$, and any background information $I$ that we may have, the \emph{likelihood function} is defined as \cite{Veitch:2008ur,Veitch:2008wd,Veitch:2009hd}
\bea
&&p(d | H_i, \vec{\theta}, I) \nn\\
& = &  \mathcal{N} \, \exp\left[ - 2 \int_{f_{\rm low}}^{f_{\rm cut}} \dee f \frac{|\tilde{d}(f) - \tilde{h}^{(i)}(f; \vec{\theta})|^2}{S_n(f)} \right], 
\end{eqnarray}
with $S_n(f)$ being the detector's one-sided noise power spectral density \cite{Maggiore:1900zz}; $\tilde{d}(f)$ is the Fourier transform of the data stream, and $\mathcal{N}$ is a normalization factor. We will take the low-frequency cutoff to be $f_{\rm low} = 40$ Hz, and $f_{\rm cut}$ is the one in Eq.~(\ref{termination}). To compute $p(d |  H_i, \vec{\theta}, I)$ we will use the method of nested sampling as implemented in GW parameter estimation by Veitch and Vecchio \cite{Veitch:2008ur,Veitch:2008wd,Veitch:2009hd}; see also \cite{Veitch:2014wba}. The \emph{evidence} is given by
\be
P(d|H_i, I) = \int \dee\vec{\theta} \, p(\vec{\theta}| I) \, p(d | H_i, \vec{\theta}, I),
\ee
with $p(\vec{\theta}| I)$ being the prior density distribution. Using Bayes's theorem, the \emph{posterior probability} of the hypothesis $H_i$ given the data $d$ is then obtained through
\be
P(H_i | d, I) = \frac{P(d | H_i, I)\,P(H_i|I)}{P(d|I)},
\ee
where $P(H_i|I)$ is the prior probability for $H_i$ before any measurement has taken place, and $P(d|I)$ is the prior probability of the data. Finally, the \emph{odds ratio} between any two hypotheses $H_i$, $H_j$ is defined as
\be
O^i_j \equiv \frac{P(H_i|d, I)}{P(H_j|d, I)} = \frac{P(H_i|I)}{P(H_j|I)} \frac{P(d|H_i, I)}{P(d|H_j, I)}.
\ee
Note that the unknown $P(d|I)$ conveniently drops out of this expression. 

The above framework can trivially be generalized to the case of multiple detections $d_1, d_2, \ldots, d_N$. Using the multiplication rule for independent random variables, one obtains
\be
{}^{(N)}O^i_j = \frac{P(H_i|I)}{P(H_j|I)} \prod_{n=1}^N \frac{P(d_n|H_i, I)}{P(d_n|H_j, I)}.
\label{combinedodds}
\ee
The probabilities $P(H_i|I)$ quantify our prior degree of belief in the hypotheses. Currently a wide range of EOS are still consistent with existing observations  \cite{Lattimer:2012nd}, including the ones that we will use in our simulations (although general theoretical considerations suggest a more restricted range; see \emph{e.g.}~\cite{Hebeler:2010jx}). In the absence of any additional information, it then makes sense to set $P(H_i|I) = P(H_j|I)$ for all $i$ and $j$. 

If ${}^{(N)}O^i_j > 1$, then the data favors the hypothesis $H_i$ over the hypothesis $H_j$. By looking at the odds ratios for all pairs of hypotheses, we arrive at an overall ranking of all the $H_1, H_2, \ldots, H_K$. We explicitly note that
\begin{enumerate}
\item Even if the true equation of state were in the set $H_i$, $i = 1, \ldots, K$, one should not necessarily expect it to end up at the top of the ranking; this is due to the effects of noise and the fact that the majority of detected sources will have low signal-to-noise ratios (SNRs). 
\item In practice, the correct equation of state will probably not be in the finite set $H_i$, $i = 1, \ldots, K$. Nevertheless, one may expect the highest-ranked hypothesis to be close to the true one.
\end{enumerate}
Here a notion of \emph{closeness} or distance in a space of functions is implied; this can be defined by \emph{e.g.}~employing the $L^{2}$ norm $||f|| = (\int |f|^2 d\mu)^{1/2}$. The integration measure $\mu$ need not be uniform in mass (\emph{i.e.}~in principle $d\mu \ne dm$), but should rather reflect the amount of information that is collected from each infinitesimal mass interval.
That is, if two functions differ significantly at a mass interval where no sources are found, but are almost equal elsewhere, then the ``distance'' between them should be small.
Here however, the set of functions $\lambda^{(i)}(m)$ that we consider are clearly distinguishable across the mass interval of interest $[1,2] \Ms$ and admit a strict ordering in terms of stiffness.

Finally, we note that it is often convenient to work with the \emph{logarithms} of the odds ratios, as we will also do here.

\subsection{Parameter estimation}
\label{sec:PE}


An obvious advantage of hypothesis ranking is that information from multiple detections can trivially be combined; see Eq.~(\ref{combinedodds}). In measuring parameters, we will want to do the same. Hence it will not do to just measure \emph{e.g.}~the tidal deformabilities $\lambda_A$, $A = 1, 2$ for each source individually, since these numbers are mass-dependent and will vary from source to source. In order to combine information across sources, one must identify observables that only depend on the equation of state and not on incidental details of the sources. In \cite{DelPozzo:2013ala}, the function $\lambda(m)$ was approximated by a Taylor expansion around some reference mass $m_0$, truncated at a suitably high order:
\be
\lambda(m) = \sum_{j = 0}^{j_{\rm max}} \frac{1}{j!} \, c_j\,\left(\frac{m - m_0}{M_\odot}\right)^j \, .
\label{lambdaTaylor}
\ee
For a given EOS, the coefficients $c_j$ will be the same for all the sources. 

A waveform model $\tilde{h}(f; \vec{\theta}, \{c_j\})$ can be constructed along the lines of Sec.~\ref{sec:waveforms}, but this time substituting the expansion (\ref{lambdaTaylor}) for $\lambda(m)$; here $\vec{\theta}$ still represents masses, spins, sky position, orientation, distance, time of coalescence, and phase at coalescence. Given a detection $d$, a \emph{posterior density function} for each of the $c_j$ can be constructed \cite{Veitch:2008ur,Veitch:2008wd,Veitch:2009hd,DelPozzo:2013ala}. For instance, in the case of $c_0$,
\be
p(c_0| d, I) = \int \dee\vec{\theta}\,\dee c_1 \ldots \dee c_{j_{\rm max}} \, p(\vec{\theta}, \{c_j\} | d, I),
\ee
where the joint posterior density function for all the parameters takes the form
\be
p(\vec{\theta}, \{c_j\} | d, I) = \frac{p(d | \vec{\theta}, \{c_j\}, I)\,p(\vec{\theta}, \{c_j\} | I)}{p(d | I)}.
\label{PDF}
\ee
Here $1/p(d | I)$ acts as a normalization factor, and $p(\vec{\theta}, \{c_j\} | I)$ is the joint prior density of all the parameters; we will assume that the latter can be written as
\be
p(\vec{\theta}, \{c_j\} | I) = p(\vec{\theta} | I)\,\prod_{j=0}^{j_{\rm max}} p(c_j|I),
\ee
where $p(\vec{\theta} | I)$ and $p(c_j|I)$ are separate prior densities for the $\vec{\theta}$ and all the $c_j$, respectively. 
Finally, the likelihood $p(d | \vec{\theta}, \{c_j\}, I)$ is given by 
\be
p(d | \vec{\theta}, \{c_j\}, I) = \mathcal{N}\, \exp \left[ - 2 \int_{f_{\rm low}}^{f_{\rm cut}} \dee f \frac{|\tilde{d}(f) - \tilde{h}(f; \vec{\theta}, \{c_j\})|^2}{S_n(f)} \right].
\ee

Given multiple detections $d_1, d_2, \ldots, d_N$, individual posterior density functions such as $p(c_0|d_n, I)$ can trivially be combined \cite{DelPozzo:2013ala}:
\be
p(c_0|d_1, d_2, \ldots, d_N) =  p(c_0|I)^{1-N}  \prod_{n = 1}^N p(c_0 | d_n),
\label{combinedPDF}
\ee
where we again assumed independence of the $d_n$ and have used Bayes's theorem. Similar expressions can of course be obtained for $p(c_j | d_1, d_2, \ldots, d_N)$, $j = 1, \ldots, j_{\rm max}$.

A choice needs to be made for the order $j_{\rm max}$ at which the Taylor expansion (\ref{lambdaTaylor}) is truncated. In \cite{DelPozzo:2013ala}, the authors chose $j_{\rm max} = 1$, under the expectation that but two coefficients will be measurable when EOS effects only enter the waveforms through the two parameters $\lambda(m_1)$, $\lambda(m_2)$. Here we will instead use a quadratic approximation to $\lambda(m)$:
\be
\lambda(m) \simeq c_0 + c_1 \left(\frac{m - m_0}{M_\odot}\right) + \frac{1}{2} c_2 \left(\frac{m - m_0}{M_\odot}\right)^2,
\label{lambdaquadratic}
\ee
with $m_0 = 1.4\,M_\odot$. Visual inspection of the $\lambda(m)$ for the EOSs considered in \emph{e.g.}~\cite{Hinderer:2009ca} (see their Fig.~2) already suggests that in the most plausible mass range for neutron stars (roughly $m \in [1, 2]\,M_\odot$), this will tend to be a good approximation, which is why we make the choice here; see also Fig.~1 for the EOSs used in this paper. As we shall see, in practice neither $c_1$ nor $c_2$ will be measurable even with a large number of sources, but $c_0$ will be.

\section{Setup of the simulations}
\label{sec:setup}

We now briefly describe how the simulations were set up. Different choices were made for the parameter distribution of the simulated signals, or injections, and for the priors on the parameters used for the data analysis.

\subsection{Injections}

For the parameters of the simulated sources, we choose astrophysically motivated distributions. Sources are placed uniformly in comoving volume in a distance range $D \in [100, 250]$ Mpc. The upper bound is approximately the angle-averaged range which Advanced LIGO is expected to reach, while the lower bound corresponds to the distance within which one can expect a detection once every two years \cite{Abadie:2010cf}; this cutoff also serves to exclude unexpectedly loud sources from our ensembles. The sky location $(\theta, \varphi)$ and orientation $(\iota, \psi)$ are both distributed uniformly on the sphere. The phase at coalescence $\varphi_c$ is taken to be  uniform in $[0, 2\pi)$. For the spins we note that observed pulsar periods and assumptions about spin-down rates lead to birth periods in the range $10  - 140$ ms \cite{Lorimer:2008se}, which corresponds to dimensionless spins $J/m^2 \lesssim 0.04$; the fastest known pulsar in a BNS system has $J/m^2 \sim 0.02$. 
The observed population is expected to spin down to much lower values of spin at the time of coalescence, but currently there is no good estimate on the spin distribution for a population of coalescences that will be observed with GW detectors. We choose to take a conservative approach and assume the spins to be small but non-negligible for the analysis: when including spins in the simulated signals, we take them to be Gaussian distributed with zero mean and a spread of $\sigma_\chi = 0.02$. 
Unless stated otherwise, the component masses are picked from a Gaussian with mean $\mu_m  = 1.35\,M_\odot$ and spread $\sigma_m = 0.05\,M_\odot$. The latter is inspired by estimates of the mass distribution of known neutron stars in BNS systems \cite{Valentim:2011vs,Kiziltan:2010ct,Kiziltan:2011mj,Kiziltan:2013oja}. 

For the EOS, we want to find out under what circumstances one will at least be able to distinguish between a stiff, a moderate, and a soft EOS. For these we choose the EOSs labeled MS1, H4, and SQM3, respectively, in Fig.~1. 

The simulated GW waveforms are added coherently to simulated data streams for Advanced LIGO detectors at Hanford, WA, and Livingston, LA, as well as an Advanced Virgo detector at Cascina, Italy. The detector noise is taken to be stationary and Gaussian, where the underlying noise curve for Advanced LIGO is the one with zero detuning of the mirror and high laser power, and for Advanced Virgo we choose a BNS-optimized curve assuming an appropriate choice for the signal recycling detuning and the signal recycling mirror transmittance; see \cite{DelPozzo:2013ala} and references therein. To realistically simulate a scenario in which BNS signals have been detected, we impose two conditions on the signals: (i) the optimal network SNR should be greater than 8 but smaller than 30, and (ii) the post-analysis log Bayes factor for signal versus noise should be greater than 32 with templates that assume point particle coalescence~\footnote{The detection efforts themselves can not take the unknown EOS effects into account, and searches are done with point particle templates.}. This is also what was done in \cite{Agathos:2013upa}, and we refer to that paper for more details.

\begin{figure*}
\label{fig:MSnospins}
\includegraphics[width=\columnwidth]{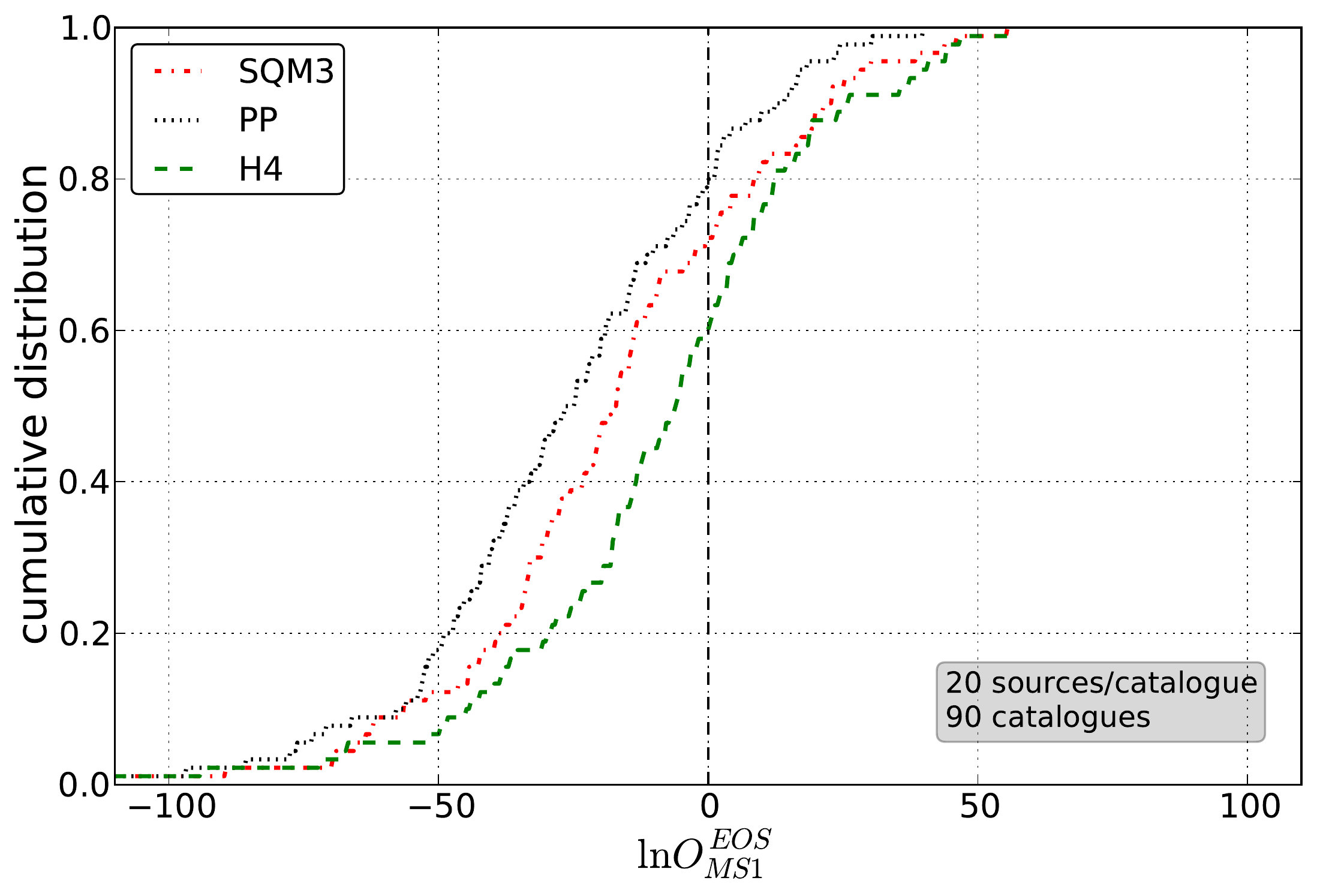} 
\includegraphics[width=\columnwidth]{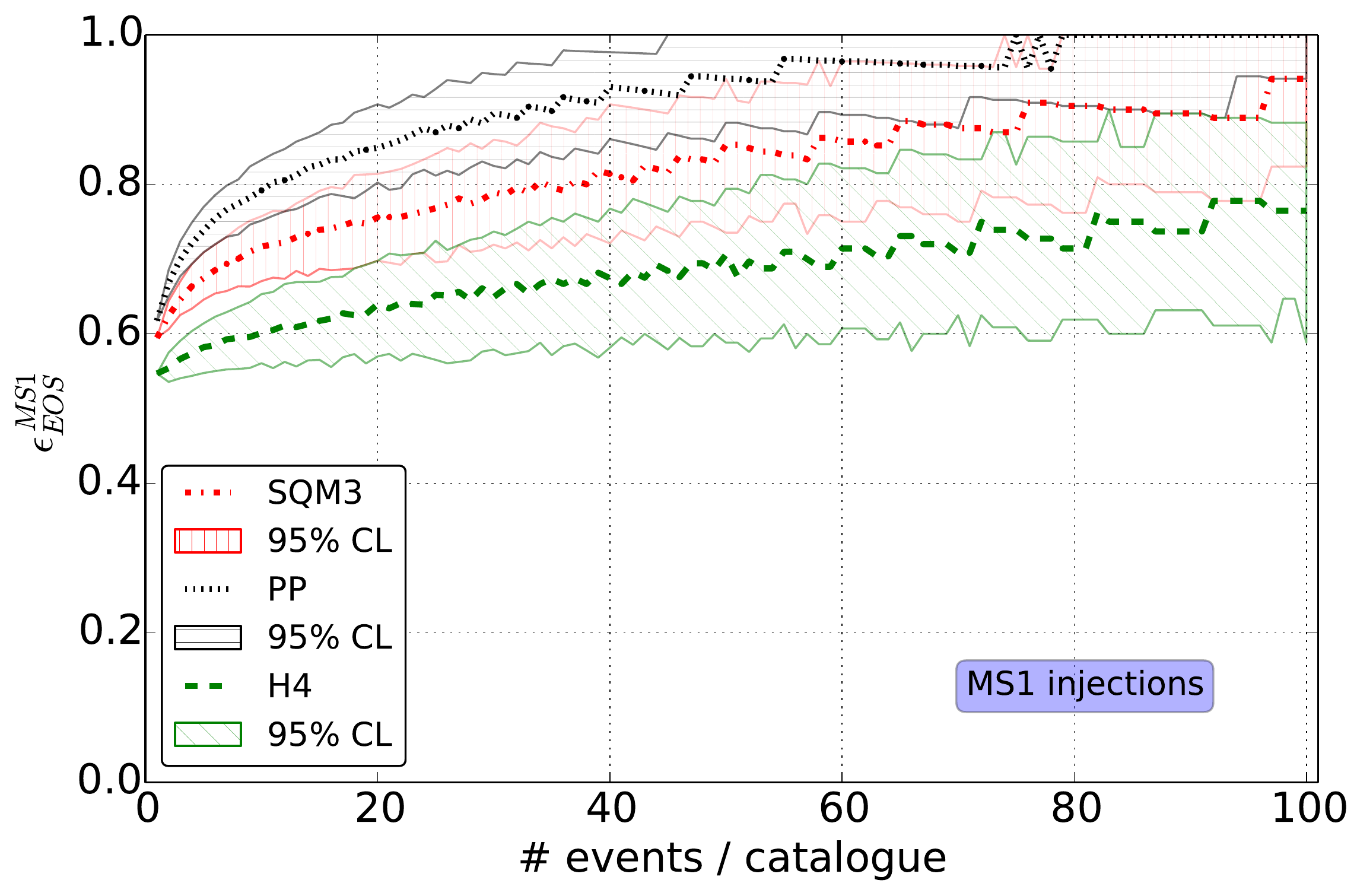} 
\caption{Hypothesis ranking in the case where the EOS of the simulated sources is MS1. As in \cite{DelPozzo:2013ala}, spins are set to zero both in the injections and the templates, and masses are distributed uniformly on the interval $[1,2]\,M_\odot$, so that any qualitative differences with previous results come from considering tidal effects to higher PN order, and terminating waveforms at contact or LSO, whichever comes sooner. Left: The cumulative distributions of the log odds ratios $\ln {}^{(20)}O^{\rm EOS}_{\rm MS1}$ for catalogs of 20 sources each, where ``EOS'' is in turn H4, SQM3, and PP. Note how the EOSs are ranked according to how dissimilar they are to the correct one: PP differs the most and is indeed the most deprecated, followed by SQM3 and H4. Right: The fraction of catalogs for which PP, SQM3, and H4 are correctly ranked lower than MS1, as a function of the number of events per catalog. What is shown are the medians and 95\% confidence intervals obtained from combining individual sources into catalogs in 1,000 different ways.}
\end{figure*}

\begin{figure*}
\label{fig:MSgaussmass}
\includegraphics[width=\columnwidth]{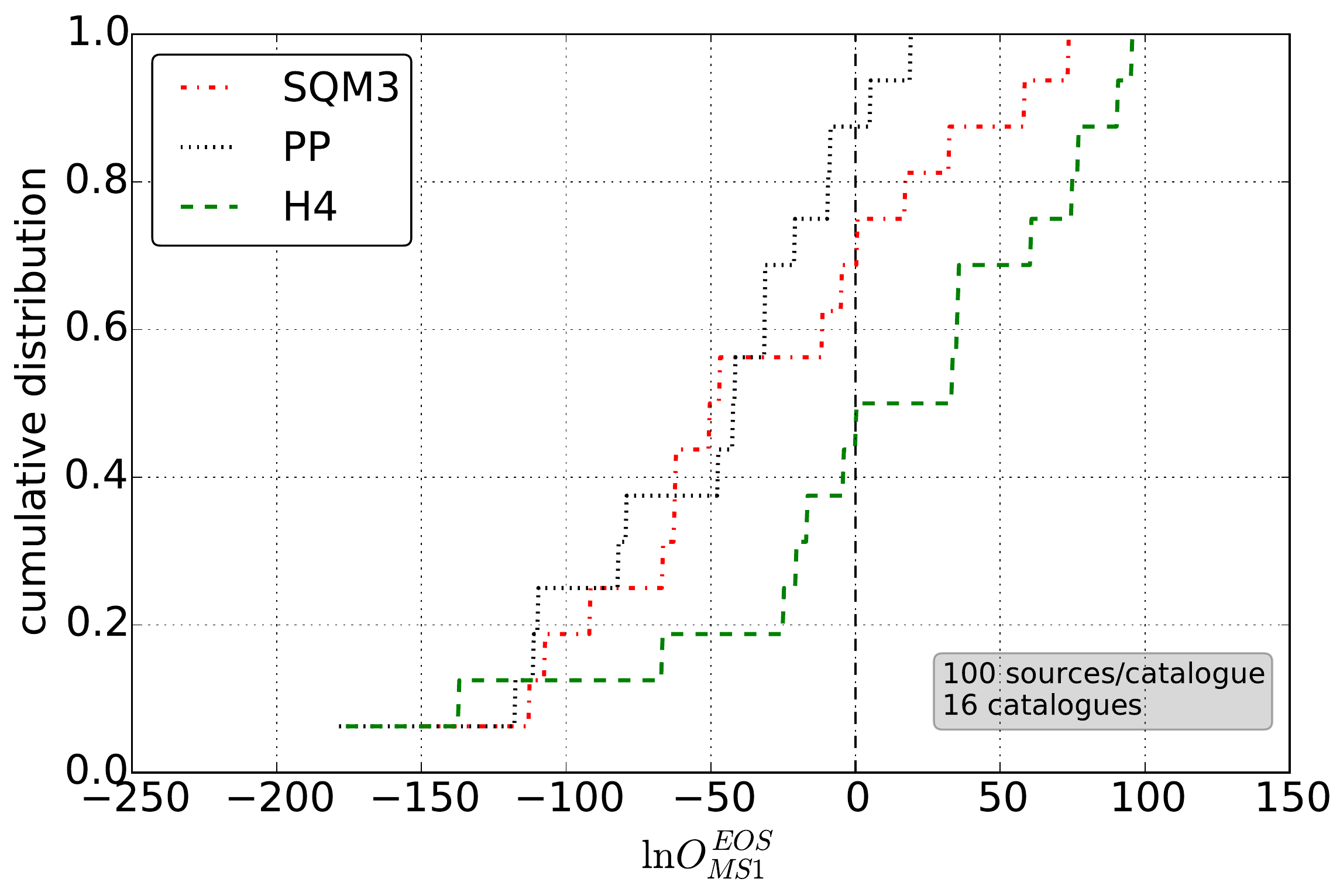} 
\includegraphics[width=\columnwidth]{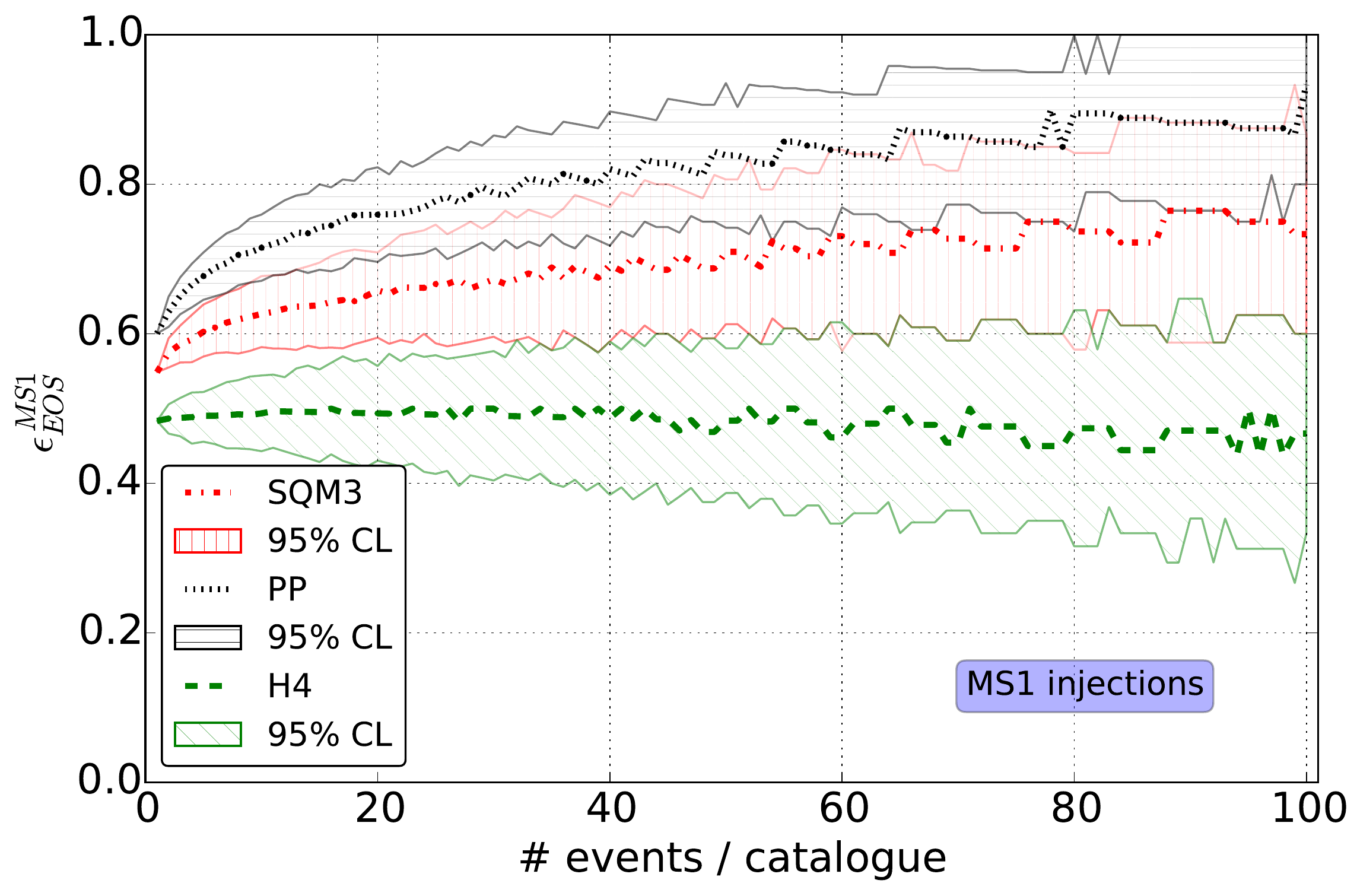} 
\caption{The same as in Fig.~5, but this time with a relatively strongly peaked Gaussian distribution for the injected component masses (while still using a flat mass prior in the analyses). \emph{To approach the discernibility of EOS seen in Fig.~5,  we now need $\mathcal{O}(100)$ sources per catalog}. Even then, H4, the EOS that is closest to the injected MS1, can not be distinguished from it.}
\end{figure*}

\subsection{Templates}

The data analysis was performed as described in Sec.~\ref{sec:methods}, with the following choices of prior distributions for the parameters. Distance was allowed to vary in the range $D \in [1, 1000]$ Mpc. The coalescence phase $\varphi_c$ was taken to be uniform on $[0, 2\pi)$. The coalescence time was allowed to vary within 100 ms of an injection. When allowing for non-zero spins in the templates, we took the priors on $\chi_1$, $\chi_2$ to be uniform on the interval $[-0.1, 0.1]$.

As we shall see, the prior density distribution for the component masses will play an important role. In principle we could take this to be the same as the mass distribution for the injections, \emph{i.e.}~Gaussian with $\mu = 1.35\,M_\odot$ and $\sigma_m = 0.05\,M_\odot$. However, we would then implicitly be assuming that the astrophysical mass distribution of neutron stars in binaries will be reliably known in the advanced detector era. At the time of writing only 9 double neutron star systems have been observed, sometimes with large error bars on the measured masses; it seems unlikely that this situation will improve dramatically in the next few years. We also note the differing results for observationally based estimates of the mass distribution in BNS systems; for example, $(\mu_m, \sigma_m) = (1.37\,M_\odot, 0.042\,M_\odot)$ in Valentim \emph{et al.}~\cite{Valentim:2011vs} and $(\mu_m, \sigma_m) = (1.33\,M_\odot, 0.13\,M_\odot)$ in Kiziltan \emph{et al.}~\cite{Kiziltan:2013oja}, the difference partially being due to the use of different subsets of the known systems based on the reliability of individual mass measurements. Finally, it is possible that due to selection biases, the distribution of masses in electromagnetically observed neutron star binaries will not be identical to the mass distribution in BNS coalescences detected by Advanced LIGO and Virgo. For these reasons, we will mostly assume a flat component mass prior with $m \in [1, 2]\,M_\odot$. However, in the Appendix we will also briefly investigate what happens if the astrophysical distribution of masses of neutron stars in binaries can be assumed known after all.

For the EOSs in the hypothesis ranking, we again consider MS1, H4, and SQM3, as well as the ``point particle'' model, denoted PP. The latter corresponds to a waveform model where $\lambda(m) \equiv 0$. 

Finally, when doing parameter estimation on the coefficients $c_0$, $c_1$, and $c_2$ in the quadratic approximation to $\lambda(m)$ as in Eq.~(\ref{lambdaquadratic}), the priors are chosen to be $c_0 \in [0, 5] \times 10^{-23} s^5$, $c_1 \in [-2.5, 0] \times 10^{-23} s^5$, and $c_2 \in [-3.7, 0] \times 10^{-23} s^5$. In the mass regime of interest, this captures all the EOSs in Fig.~2 of \cite{Hinderer:2009ca}. 

\section{Results}
\label{sec:results}

Let us now present the results of our simulations, first for the hypothesis ranking described in Sec.~\ref{sec:MS} and then for parameter estimation as explained in Sec.~\ref{sec:PE}.


\subsection{Hypothesis ranking}

A first estimate of how well one will be able to determine the EOS using hypothesis ranking was presented in our earlier paper \cite{DelPozzo:2013ala}. In that work, only tidal effects up to 1PN rather than 2.5PN were taken into account, quadrupole-monopole contributions were disregarded, waveforms were terminated at LSO instead of the frequency $f_{\rm cut}$ of Eq.~(\ref{termination}), and spins were set to zero both in the injections and in the template waveforms. Additionally, the component masses were taken to be distributed uniformly in the interval $[1,2]\,M_\odot$ rather than according to a Gaussian with mean $\mu_m = 1.35\,M_\odot$ and spread $\sigma_m = 0.05\,M_\odot$. 

Ideally one would like to look at the impact of each of these effects individually. However, the simulations presented in this paper are computationally expensive if one wants to have good statistics. For this reason, we proceed as follows:
\begin{itemize}
\item First we set the spins to zero both in injections and templates (so that the quadrupole-monopole effect is not present), but we take tidal effects to 2.5PN and terminate the waveforms at the minimum of contact frequency and LSO frequency. We generate results for injected component masses distributed uniformly in $[1,2]\,M_\odot$, and then for component masses following a Gaussian with $\mu_m = 1.35\,M_\odot$ and $\sigma_m = 0.05\,M_\odot$; however, in both cases the mass prior in our Bayesian analysis is taken to be uniform on $[1,2]\,M_\odot$. Again because of computational cost, we only make this comparison for the case where the EOS in the signals is MS1, \emph{i.e.}~the stiffest equation of state considered in this paper. 
\item Next we specialize to the more astrophysically motivated Gaussian distribution for the component masses (still keeping a uniform prior in the analysis), and we also switch on spins. In the injections, we let the latter be Gaussian distributed with zero mean and $\sigma_\chi = 0.02$, while in the templates we let the prior on the spins be uniform on the interval $[-0.1, 0.1]$, to reflect the ignorance about spins we will in practice have. Since in this case we are including all the astrophysical effects considered in this paper, we generate results not only for MS1, but for H4 and SQM3 as well.
\end{itemize}

\subsubsection*{1. Zero spins; flat versus Gaussian distribution of component masses}

First we consider the case of zero spins in injections and templates, and component masses are distributed uniformly on the interval $[1,2]\,M_\odot$. Results are shown in Fig.~5. We let the injections have MS1 as their EOS, and as in \cite{DelPozzo:2013ala}, we compute the log odds ratios $\ln {}^{(20)}O^{\rm EOS}_{\rm MS1}$ for catalogs of 20 sources each, where, in turn, ``EOS'' stands for PP, SQM3, and H4. Examples of the cumulative distributions of these log odds ratios are shown in the left panel of the figure. In the absence of detector noise, one would have $\ln {}^{(20)}O^{\rm EOS}_{\rm MS1} < 0$ in all three cases, since any EOS different from the correct one (MS1) would be deprecated. What we see is that $\ln {}^{(20)}O^{\rm PP}_{\rm MS1} < 0$ for about 80\% of the catalogs, while $\ln {}^{(20)}O^{\rm H4}_{\rm MS1} < 0$ in about 60\% of the cases. Note that H4 is the most similar to MS1, followed by SQM3 and PP; and indeed, the log odds ratios obtained tend to correctly rank the EOSs in this way. This is similar to what one sees in the top right panel of Fig.~2 of \cite{DelPozzo:2013ala}. However, despite the fact that in the present work we take tidal effects to much higher order, the left tails of the cumulative log odds ratio distributions stretch to less negative values. This can be explained by the different termination of the waveforms, which for the EOSs and mass distributions we consider tends to be at contact rather than LSO (see Fig.~3). For a typical system with component masses $(1.35, 1.35)\,M_\odot$ and equation of state MS1, the termination frequency is $f_{\rm contact} = 1222$ Hz whereas $f_{\rm LSO} = 1629$ Hz, so that the signal contains less information on tidal effects (which manifest themselves at high frequency) than in \cite{DelPozzo:2013ala}.
Indeed, as can be seen in Fig.~4, the higher-order tidal effects due to their alternating signs do not significantly change the number of cycles in the phase, though they will add some structure because they come with different powers of $v$; on the other hand, termination at contact seems to have a much stronger effect, cutting the tidal phase short (in this example by roughly 15 rad).
The QM effect is much weaker and is not expected to give a significant contribution to the inference.

The right panel of Fig.~5 shows the fraction $\epsilon^{\rm MS1}_{\rm EOS}$ of catalogs for which MS1 is ranked higher than, respectively, H4, SQM3 and PP (\emph{i.e.}~$\ln  {}^{(20)}O^{\rm EOS}_{\rm MS1} < 0$ where ``EOS'' is, in turn, PP, SQM3, and H4), as a function of the number of sources per catalog. Also shown are 95\% intervals on $\epsilon^{\rm MS1}_{\rm EOS}$ obtained from combining 1,800 individual sources into catalogs in 1,000 different ways. We see the same trend as in the left panel: H4, being the most similar to MS1, is ranked below MS1 the least often, and PP, being the most dissimilar, the most often. We note that in going to a higher number of sources per catalog, we start experiencing small number statistics; at 100 sources per catalog only 18 independent catalogs can be composed.

Next, in Fig.~6 we look at the case where the spins are still zero in injections and templates, but the injected masses are distributed according to a Gaussian with $\mu_m = 1.35\,M_\odot$ and $\sigma_m = 0.05\,M_\odot$. Unlike in Fig.~5, in the left panel showing the cumulative distributions of the log odds ratios, \emph{we now consider catalogs of 100 sources each}, which turns out to be necessary to approach the discriminatory power we had with a uniform mass distribution. Even then, H4, the EOS that most closely resembles the injected MS1, is not distinguishable from it: the probability that MS1 gets ranked above H4 is approximately the same as the probability that H4 ends up above MS1.

\subsubsection*{2. Gaussian mass distribution, non-zero spins}

\begin{figure*}
\label{fig:MSs02}
\includegraphics[width=\columnwidth]{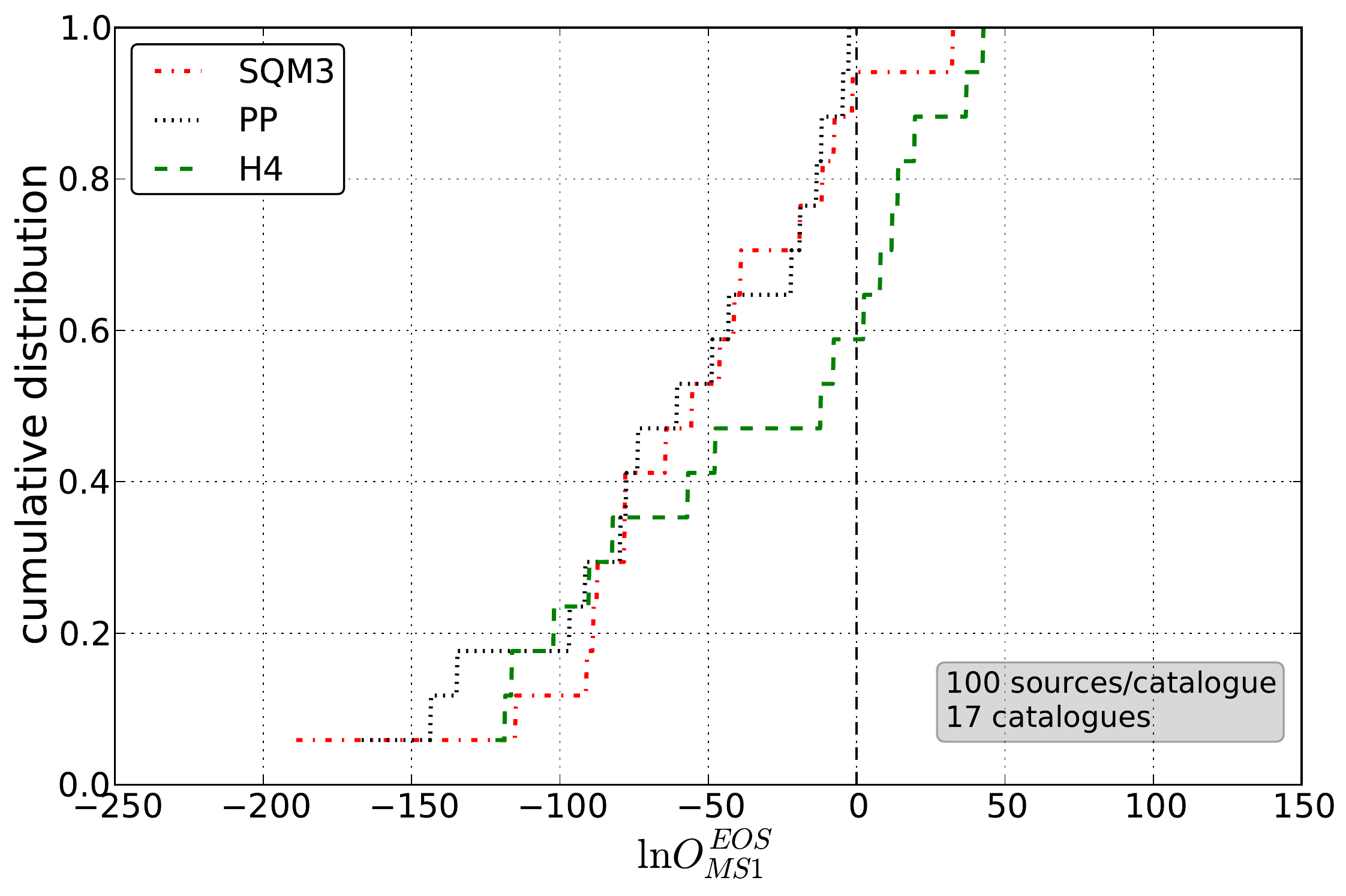} 
\includegraphics[width=\columnwidth]{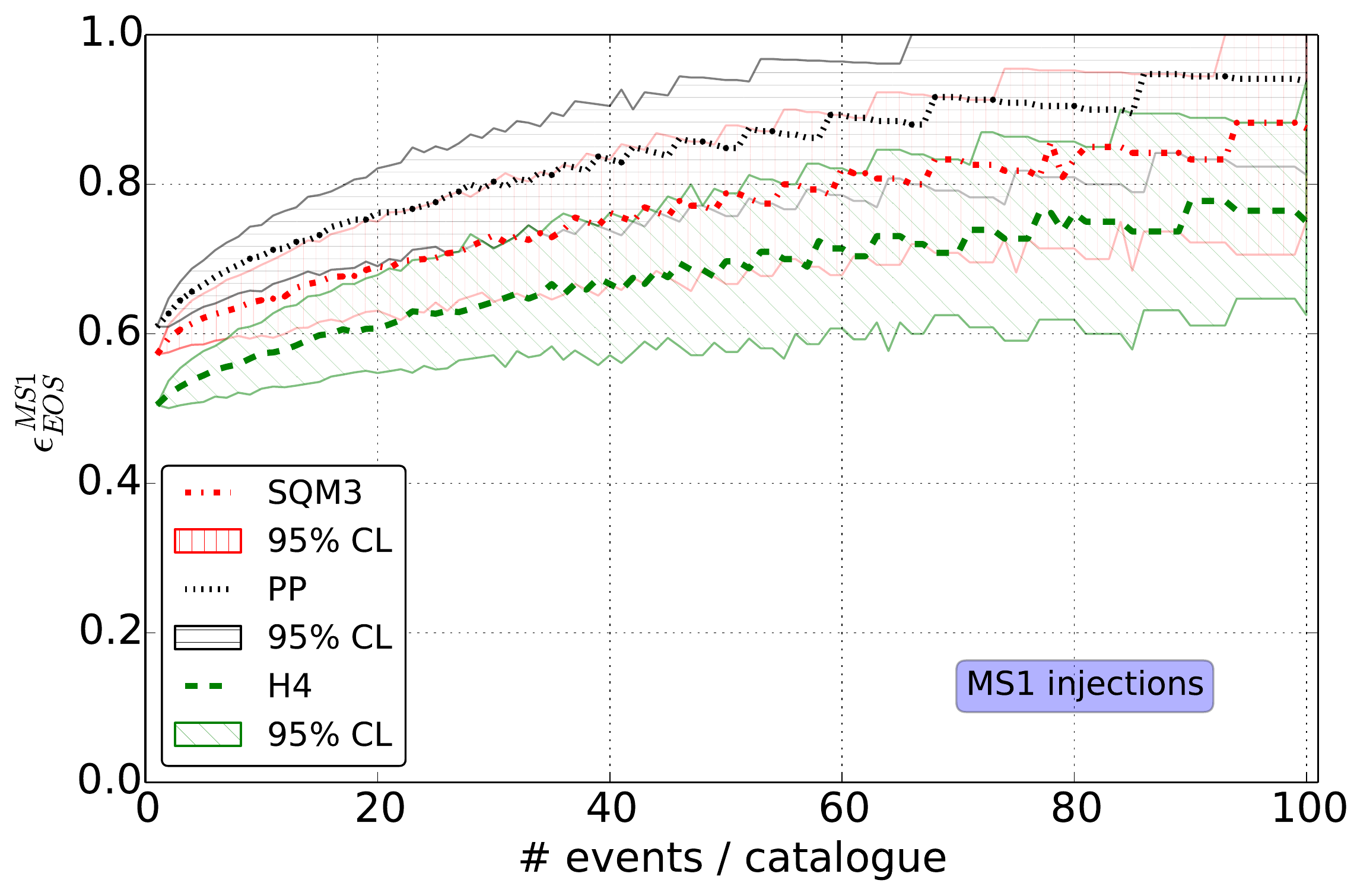}  
\\
\includegraphics[width=\columnwidth]{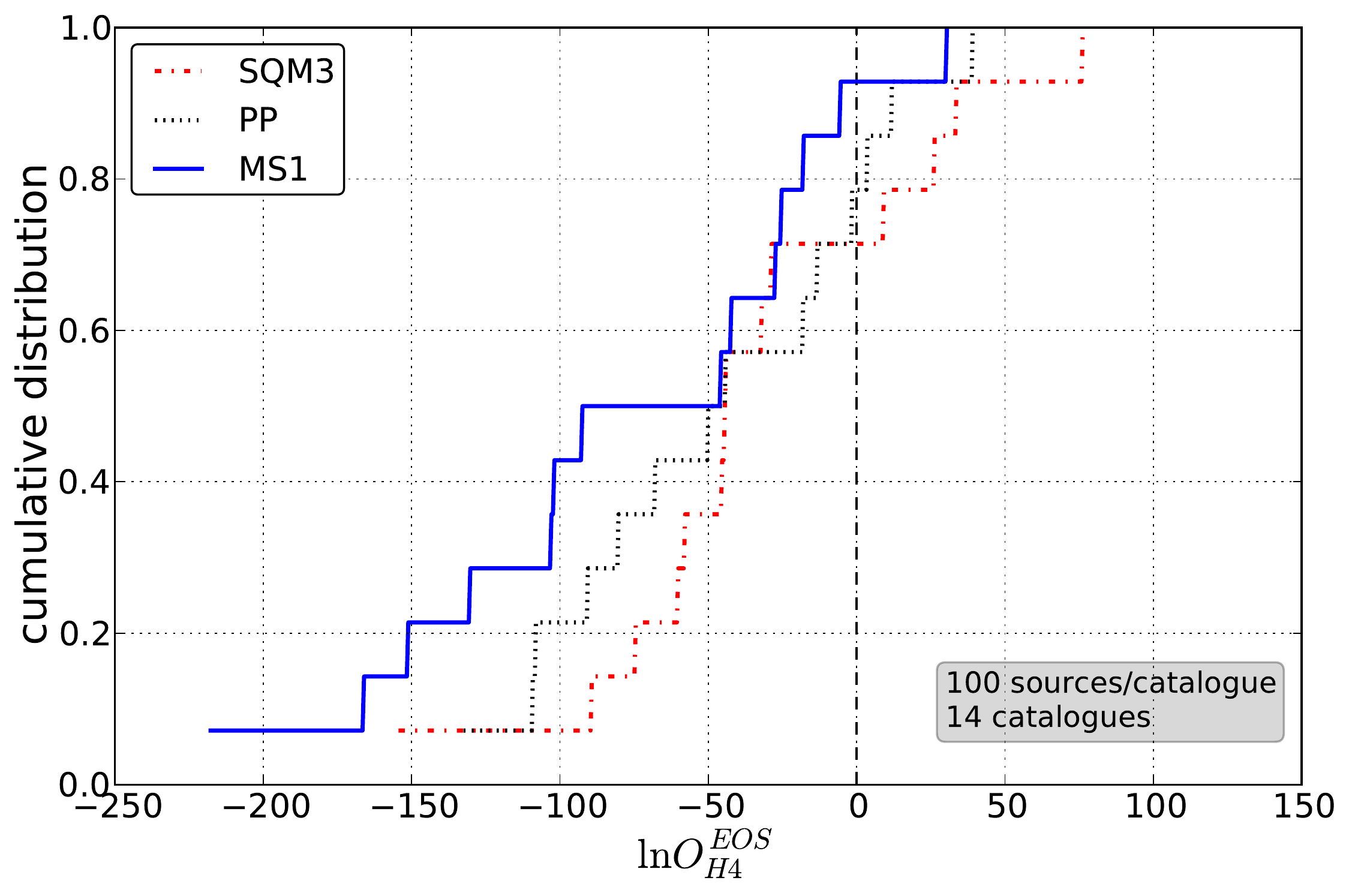} 
\includegraphics[width=\columnwidth]{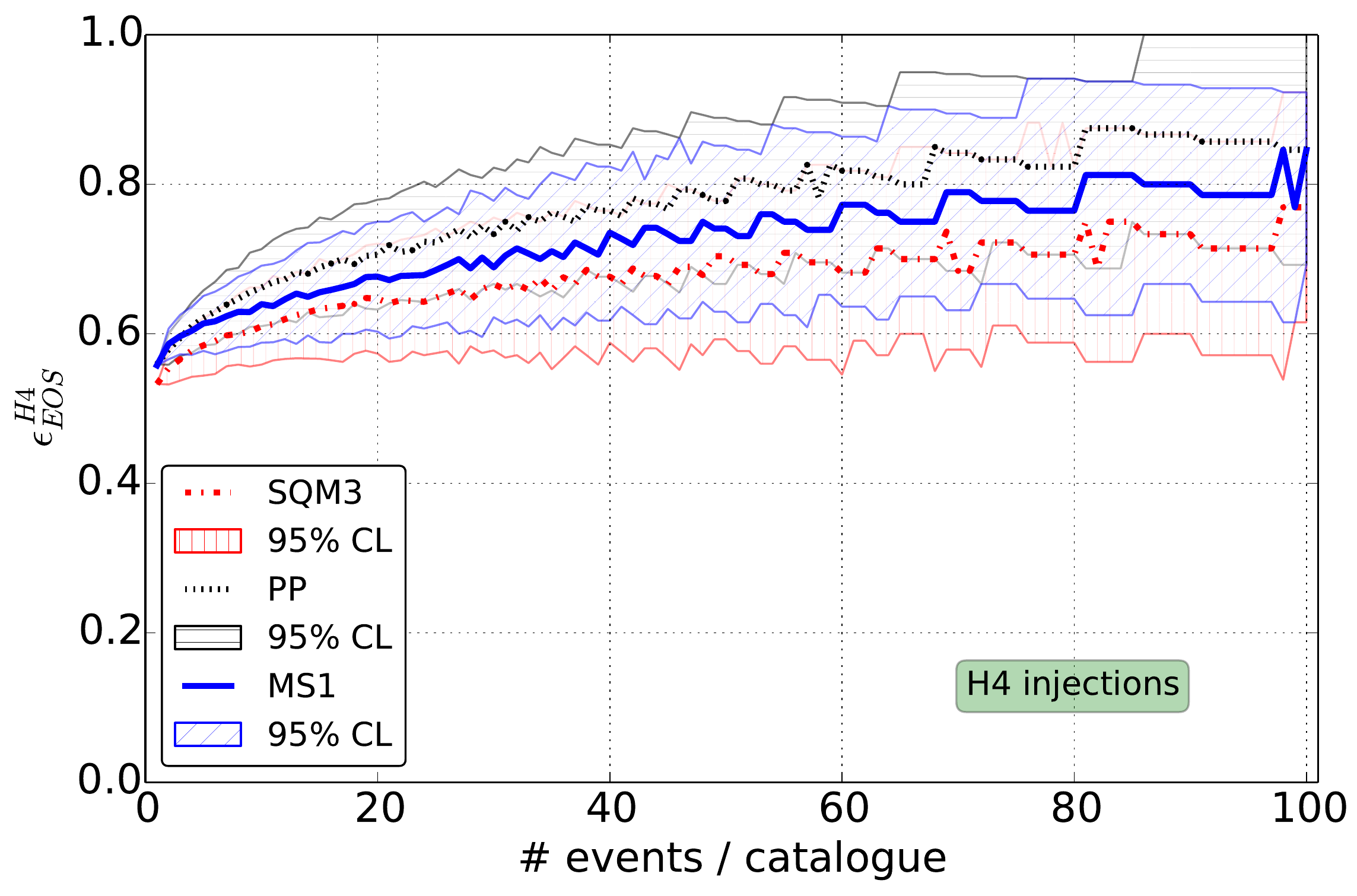}  
\\
\includegraphics[width=\columnwidth]{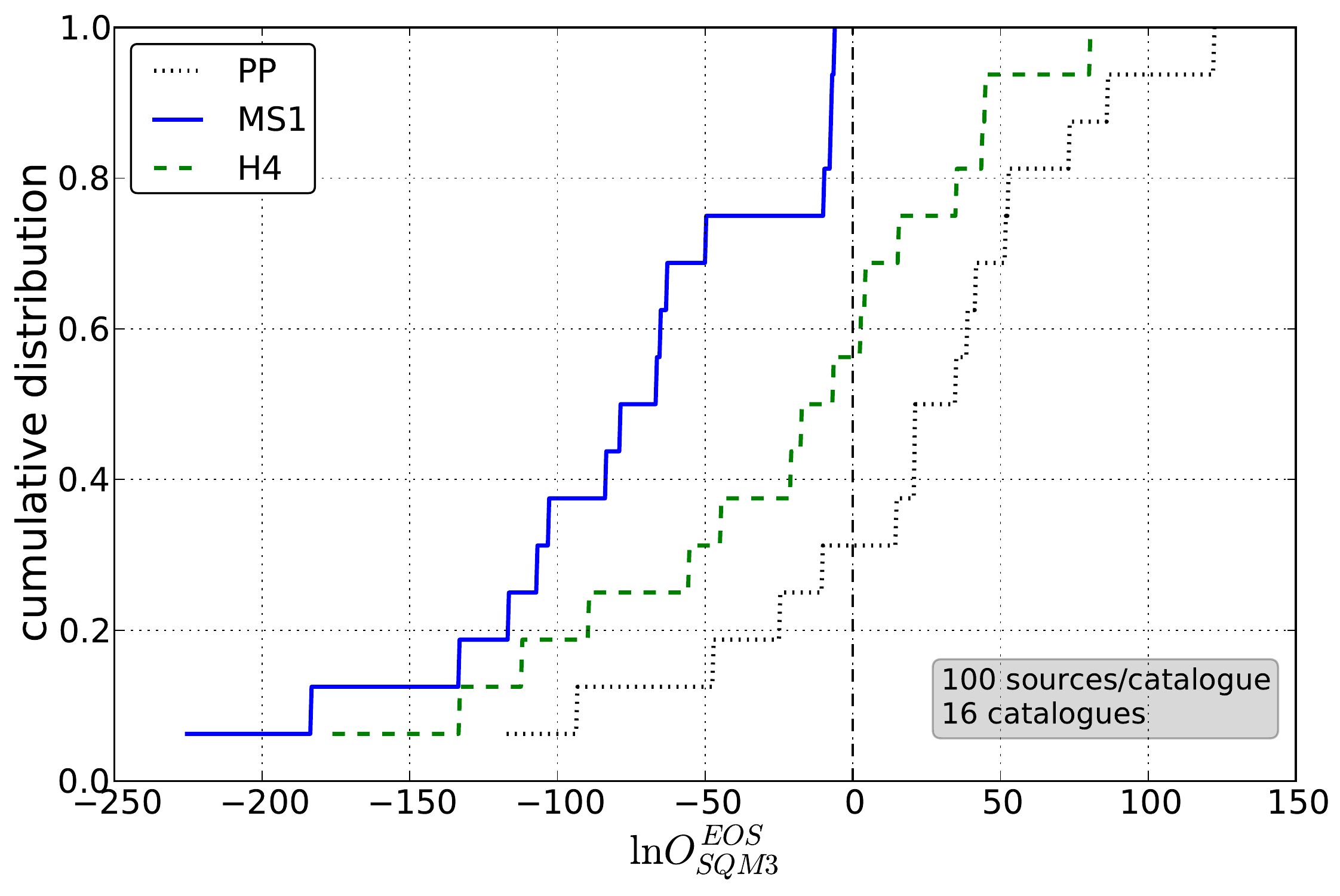} 
\includegraphics[width=\columnwidth]{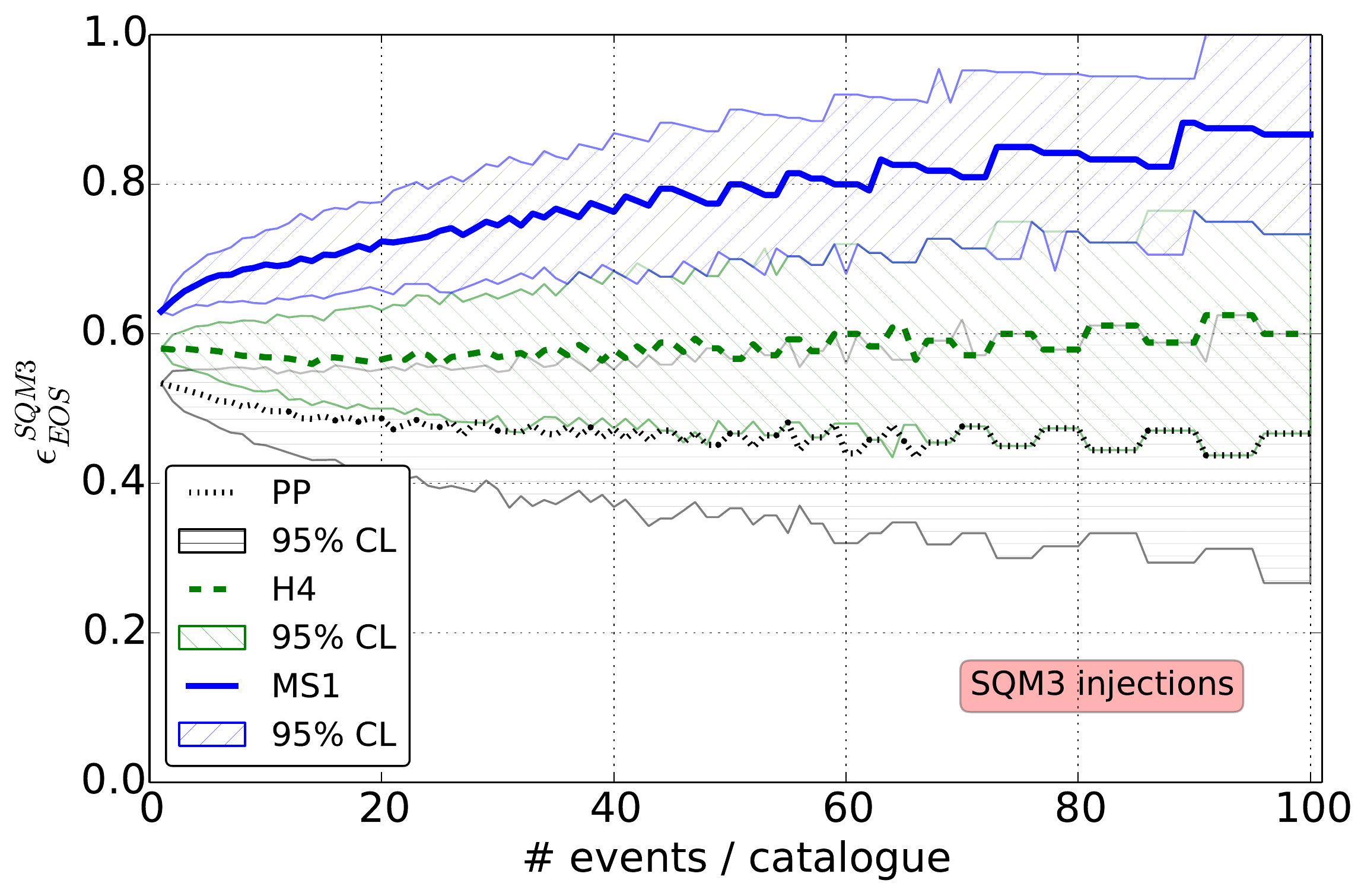}  
\caption{The same as in Fig.~6, except that simulated sources have (anti-)aligned spins sampled from a Gaussian distribution  centered at zero and with $\sigma_\chi = 0.02$. However, the prior on spins used in the recovery is uniform on the interval $[-0.1, 0.1]$. Left panels: Examples of cumulative distributions of log odds ratios for the injected EOS versus the other ones considered, for catalogs of 100 sources each. From top to bottom the injected EOS is MS1, H4, and SQM3, respectively. Right panels: The fraction of catalogs for which the correct EOS is ranked higher than each of the others in turn, as a function of the catalog size. Here too, medians and 95\% confidence intervals are shown, obtained from combining sources into catalogs in 1,000 different ways.}
\end{figure*}

We now specialize to the astrophysically better motivated Gaussian distribution for the injected component masses (but sticking to a uniform mass prior in our analyses), and we switch on spins $\chi_A$, $A = 1, 2$. In the injections, the spins are Gaussian distributed with zero mean and $\sigma_\chi = 0.02$, while in the templates, the $\chi_A$ have priors that are uniform on the interval $[-0.1, 0.1]$. This time we give results for injections where the EOS is MS1, H4, and SQM3, respectively.

In the left panels of Fig.~7 we see examples of cumulative distributions of $\ln {}^{(100)}\mathcal{O}^{\rm EOS}_{\rm inj}$ for catalogs of 100 sources each, where ``inj'' is the injected equation of state, while ``EOS'' is, in turn, taken to be each of the other three EOSs considered in this paper. From top to bottom, the injections follow MS1, H4, and SQM3, respectively.  

In the right panels of Fig.~7 we again vary the number of sources per catalog, and show the fraction $\epsilon^{\rm inj}_{\rm EOS}$ of times that the injected equation of state is ranked higher than each of the other three EOSs in turn. For a given number of sources per catalog, we combine individual sources into catalogs in many different ways and look at the medians and 95\% confidence intervals of the $\epsilon^{\rm inj}_{\rm EOS}$.

Let us first compare the results for MS1 (top panels in Fig.~7) with the ones for Gaussian distributed masses but zero spins in injections and templates (Fig.~6). Looking at the $\epsilon^{\rm MS1}_{\rm EOS}$, we infer that EOSs again tend to be ranked correctly according to ``stiffness'' and similarity to MS1, and we even see some improvement in the discernibility of H4 from MS1, especially as the number of sources per catalog goes to 100. 

For H4 injections (middle panels in Fig.~7), the medians of $\epsilon^{\rm H4}_{\rm EOS}$ are still ordered, with the median of $\epsilon^{\rm H4}_{\rm PP}$ staying above that of $\epsilon^{\rm H4}_{\rm MS1}$, which in turn trumps $\epsilon^{\rm H4}_{\rm SQM3}$. However, H4 being in between MS1 and SQM3 in stiffness (see Fig.~1), the 95\% uncertainty intervals of the $\epsilon^{\rm H4}_{\rm EOS}$ show considerable overlap; although H4 is ranked above each of the other EOSs reasonably frequently, the internal ranking is less clear. 

Finally, for SQM3 (bottom panels), this being the softest EOS other than the PP model, the stiff MS1 tends to be deprecated reasonably strongly, but it is hard to distinguish SQM3 from either H4 or PP. 

\subsection{Parameter estimation}

We now turn to the data analysis setup described in Sec.~\ref{sec:PE}. Here the templates used for the recovery do not have a fixed tidal deformability function $\lambda(m)$; rather, it is modeled by a quadratic polynomial as in Eq.~(\ref{lambdaquadratic}), where the coefficients $c_0$, $c_1$, $c_2$ are now free parameters to be estimated, on top of all the usual ones (masses, spins if applicable, time and phase at coalescence, sky position, orientation, and distance). To the extent that the quadratic approximation can capture the EOS in the signal in the relevant mass range, in the measurement process we can assume $c_0$, $c_1$, and $c_2$ to have fixed values, so that their posterior densities can be combined across sources as in Eq.~(\ref{combinedPDF}). 

In our earlier paper \cite{DelPozzo:2013ala}, where only a linear approximation to $\lambda(m)$ was used, it was found that only the zeroth-order coefficient was measurable. The quadratic approximation used in the present paper should allow for a better fit, but here too, it turns out that only the leading-order coefficient $c_0$ can be measured with any kind of accuracy.  Thus, unlike with hypothesis ranking, in practice only a single number pertaining to the EOS is being extracted from the data. Nevertheless, one has $c_0 = \lambda(m_0)$, with $m_0$ some fixed reference mass (which we will take to be $1.4\,M_\odot$), and as can be seen in Fig.~2 of the paper by Hinderer \emph{et al.} \cite{Hinderer:2009ca}, which shows nearly 20 different predictions for $\lambda(m)$, valuable information could be gleaned from just that one number. 

As before, we consider the following cases:
\begin{itemize}
\item Spins are zero both in injections and templates, and we compare results for an injected mass distribution that is uniform on $[1,2]\,M_\odot$ with what one gets with a Gaussian mass distribution that has $\mu_m  = 1.35\,M_\odot$ and $\sigma_m = 0.05\,M_\odot$. However, for the templates we do not assume knowledge of the astrophysical mass distribution, sticking to a uniform mass prior on $[1,2]\,M_\odot$.
\item Next we specialize to the Gaussian injected mass distribution, and switch on spins. In the injected waveforms, the latter are drawn from Gaussian distributions with zero mean and $\sigma_\chi = 0.02$, while in the templates the priors for the spins are uniform on $[-0.1, 0.1]$.
\end{itemize}

We stress again that for analysis purposes we will not assume knowledge of the astrophysical mass distribution, and we will use a prior on the component masses that is uniform on the interval $[1,2]\,M_\odot$. As we shall see, significant biases will appear in the estimation of $c_0$. These can be traced back to this flat prior. As demonstrated in the Appendix, if we had exact knowledge of the astrophysical mass distribution and could use that as a prior instead, the biases would go away.  

\subsubsection*{1. Zero spins; flat versus Gaussian distribution of component masses}

Let us start with the case of zero spins, and a uniform mass distribution. Fig.~8 shows the evolution of the medians and 95\% confidence intervals in the measurement of $c_0$ as information from an increasing number of detected sources is combined, the injected EOS in turn being MS1, H4, and SQM3. We see that a clean separation between posterior densities occurs after $\sim 50$ sources have become available, and uncertainties of $\sim 10\%$ are reached as the number of detections goes towards 100. This can be compared with Fig.~1 of our earlier paper \cite{DelPozzo:2013ala}, where the separation also happens around $\sim 50$ sources, but $\sim 10\%$ errors are arrived at somewhat sooner than here. We recall that in that work, tidal effects were only taken to 1PN order; on the other hand, waveforms were terminated at the LSO frequency rather than at the minimum of the LSO and contact frequencies. The earlier termination of signal waveforms in the present paper leads to a smaller number of cycles, and somewhat less information about the EOS is available. 

\begin{figure}
\label{fig:PEs0205} 
\includegraphics[width=\columnwidth]{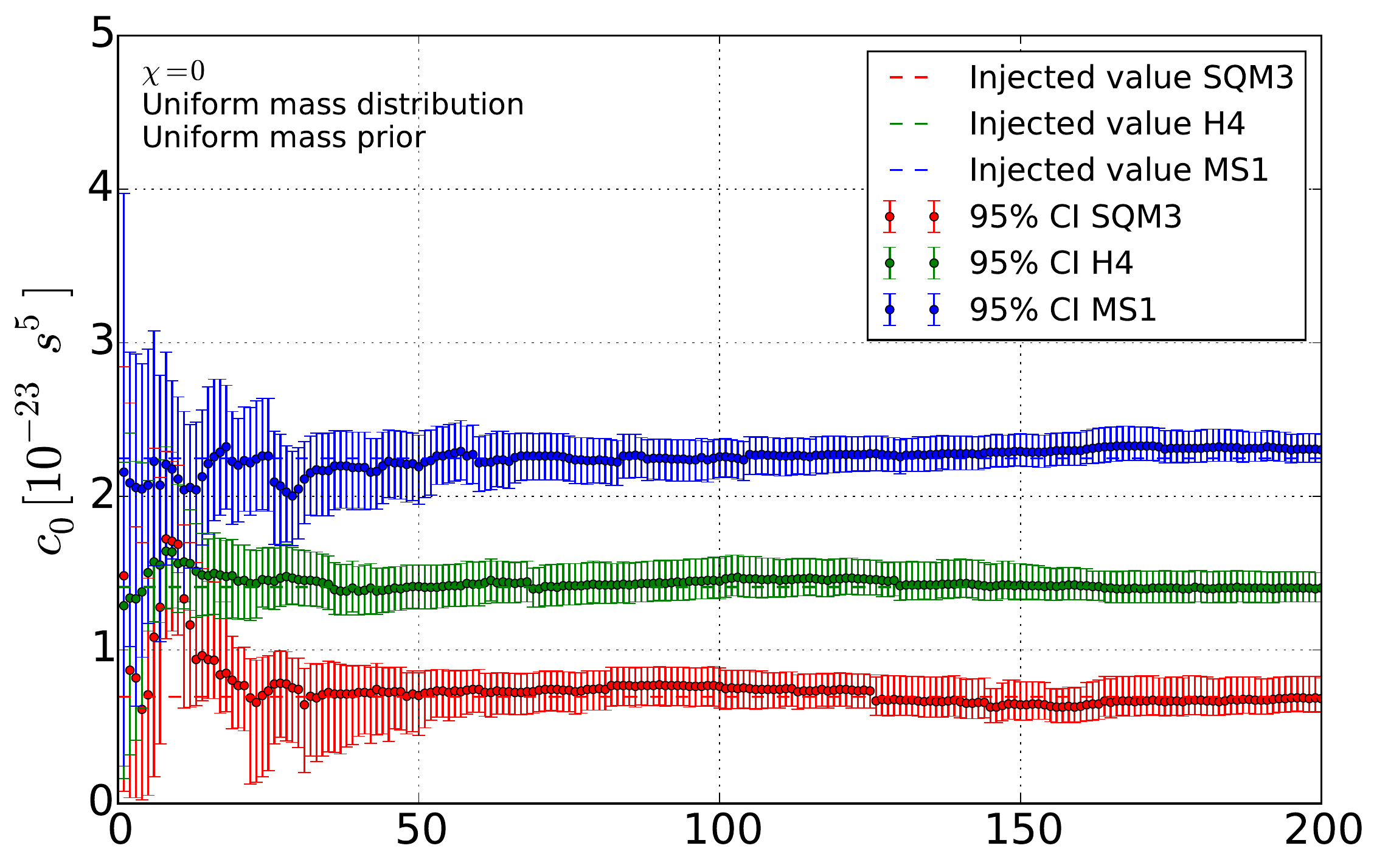} 
\\
\caption{Evolution of the medians and 95\% confidence intervals in the measurement of $c_0 = \lambda(m_0)$, the tidal deformability at the reference mass $m_0 = 1.35\,M_\odot$, for the cases where the injected EOS is MS1, H4, or SQM3. Both in the injections and the templates, spins are set to zero, and the injected mass distribution is uniform on the interval $[1, 2]\,M_\odot$.}
\end{figure}

In Fig.~9, we show results for zero spins, and this time a Gaussian distribution for the injected component masses. A good separation between MS1, H4, and SQM3 does not occur until $\sim 150$ sources have become available, and large systematic biases appear. As explained below, this is related to the continued use of a flat prior on the component masses, a distribution which now has a significant mismatch with the astrophysical one. The effect of the mass prior is further investigated in the Appendix.

\begin{figure}
\label{fig:PE1} 
\includegraphics[width=\columnwidth]{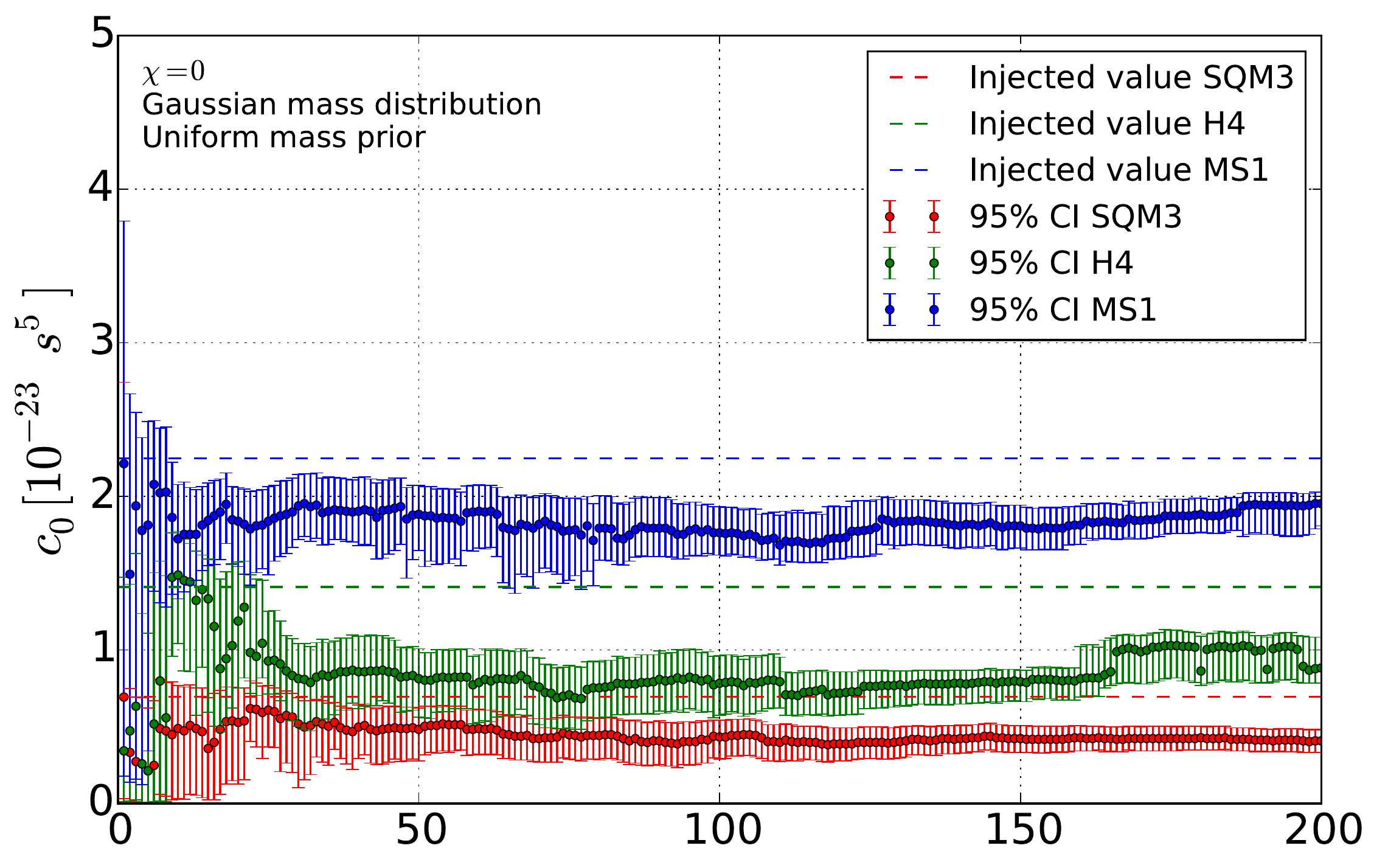} 
\\
\caption{The same as in Fig.~8, but this time the signals have component masses drawn from a strongly peaked Gaussian distribution; on the other hand, the prior distribution for the masses used in the analysis of the data is still taken to be uniform on $[1,2]\,M_\odot$. Note how large systematic errors appear. The effect of the mass prior is further investigated in the Appendix.} 
\end{figure}

\subsubsection*{2. Gaussian mass distribution, non-zero spins}

We now focus on the case of a Gaussian distribution for the injected component masses, and also switch on spins, which are drawn from a Gaussian distribution with zero mean and $\sigma_\chi = 0.02$. We also allow for spins in the template waveforms, with a prior distribution that is uniform on $[-0.1, 0.1]$, to reflect the ignorance of the true distribution of spins that we will have in reality. The results are shown in Fig.~10. As in the non-spinning case with the same injected mass distribution, there are systematic biases. Having to estimate the spins as additional parameters also increases the statistical errors, because of the larger dimensionality of the parameter space to be probed by the sampling algorithm.

\begin{figure}
\label{fig:PEs3} 
\includegraphics[width=\columnwidth]{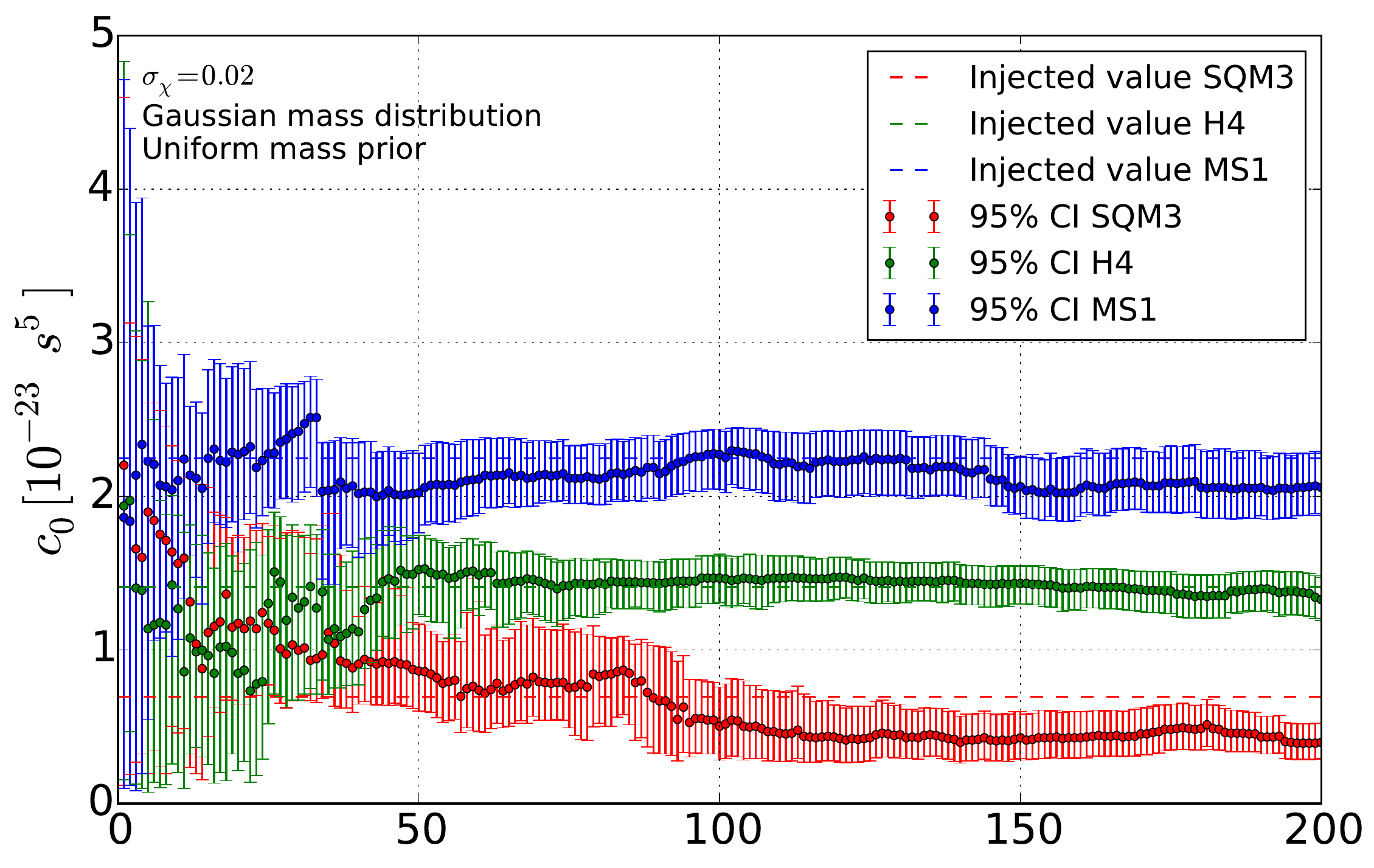} 
\\
\caption{The same as in Fig.~9, but now the signals not only have Gaussian distributed masses, but non-zero spins as well. Systematic errors remain, and statistical errors have increased due to the larger parameter space that needs to be probed.} 
\end{figure}

Finally, we mention that the higher-order coefficients $c_1$ and $c_2$ are essentially unmeasurable in all the cases we considered (with or without a Gaussian mass distribution or spins); even with 100 sources, the posteriors are not significantly different from the priors.

\section{Discussion}
\label{sec:discussion}

We have revisited the question of how well the equation of state of neutron stars can be measured with observations of binary neutron star inspirals using Advanced Virgo and Advanced LIGO. Our starting points were the Bayesian model selection and parameter estimation frameworks introduced in our earlier paper \cite{DelPozzo:2013ala}. Given a set of hypotheses associated with a list of different EOSs one can calculate the odds ratios for all pairs in the set, which provides a ranking in which EOSs that are more similar to the underlying one will tend to come out near the top, whereas EOSs that differ from it signficantly will get deprecated. Another way to gain information about the EOS from multiple sources is to model the tidal deformability $\lambda(m)$ as a series expansion in $(m - m_0)/M_\odot$ (with $m_0$ some reference mass), which is truncated at some suitable order. Since the coefficients in such an expansion are source-independent, their posterior density distributions can be combined. For the EOS we considered a ``stiff'' (MS1), ``moderate'' (H4), and ``soft'' (SQM3) equation of state, as well as the point particle model (PP). In \cite{DelPozzo:2013ala} it was found that for $m_0 = 1.4\,M_\odot$, the deformability $\lambda(m_0)$ could be determined with $\sim 10\%$ accuracy by combining information from $\mathcal{O}(20)$ sources. This was confirmed in recent work by Lackey and Wade \cite{Lackey:2014fwa}, who used a qualitatively similar waveform model as in \cite{DelPozzo:2013ala} but implemented a more physical parametrization of the EOS in terms of piecewise polytropes.


We have significantly extended our earlier study \cite{DelPozzo:2013ala}, not only by expanding the number of simulated BNS sources, but also by incorporating as much of the relevant astrophysics as has been analytically modeled, such as tidal effects to the highest known order \cite{Damour:2012yf}, neutron star spins, the quadrupole-monopole interaction \cite{Poisson:1997ha,Laarakkers:1997hb}, the impact of possible early waveform termination due to the finite radii of the neutron stars, and a strongly peaked Gaussian distribution of the component masses \cite{Valentim:2011vs,Kiziltan:2010ct,Kiziltan:2011mj,Kiziltan:2013oja}. 

In order to separate the impact of spins from the other effects, we first set spins to zero both in injections and templates (in which case the QM effect is also absent) while retaining the tidal effects as well as the potentially earlier termination of the waveform, and looked at hypothesis ranking for MS1 injections. When choosing a wide, uniform distribution for the component masses, we saw that, as in \cite{DelPozzo:2013ala}, EOSs tend to be ordered correctly according to stiffness and similarity to the true EOS. On the other hand, the log odds ratios between the incorrect and correct EOSs seemed to stretch to less negative values, presumably because of early waveform termination. Nevertheless (and again as in \cite{DelPozzo:2013ala}), hypothesis ranking worked well with catalogs of $\mathcal{O}(20)$ detected sources. The picture changed dramatically when the injected mass distribution was taken to be a strongly peaked Gaussian while keeping the mass prior to be uniform and wide as before. In that case $\gtrsim 100$ detections were needed to approach the discernibility of EOS seen in earlier work. Next we focused on a Gaussian distribution for the masses, and switched on spins. At least for MS1 injections, this turned out not to have a significant additional detrimental effect on our ability to distinguish between the EOSs. For H4, being in between MS1 and SQM3 in terms of stiffness, we saw that the correct EOS got ranked above the others a reasonable fraction of the time, but the internal ordering became less clear. Finally, for SQM3, even with catalogs of 100 sources only MS1 could be distinguished from the injected EOS reasonably well, but not H4 or PP.

We also looked at parameter estimation for the coefficients in a series expansion of $\lambda(m)$ in the small quantity $(m - m_0)/M_\odot$, truncated at some suitable order. Contrary to our earlier work we used a quadratic rather than a linearized approximation; nevertheless we found that, here too, only the leading-order coefficient is measurable. When the signals have a strongly peaked Gaussian mass distribution rather than a flat one, again keeping the wide, flat mass prior, systematic errors are introduced. Switching on spins as additional parameters also increases the statistical errors.

In the Appendix we investigated the effect on parameter estimation of the prior on the masses. We found that, if we can assume to have exact knowledge of the astrophysical distribution of the source masses so that it can be used as the prior distribution, the biases in the estimation of $c_0$ largely disappear. Recent estimates for this distribution \cite{Valentim:2011vs,Kiziltan:2010ct,Kiziltan:2011mj,Kiziltan:2013oja} are based on a rather small number of observed BNS systems and show dependence on the methodology used; hence it seems that we cannot confidently claim to have detailed knowledge. One could consider supplementing the existing information with component mass measurements from the gravitational wave signals themselves, but as is well known, these will come with large uncertainties \cite{Veitch:2014wba}; moreover, due to selection biases, the distribution of masses in electromagnetically observed neutron star binaries may differ from the mass distribution in BNS coalescences seen by gravitational wave detectors. 
A more extensive investigation of the effect of the prior distribution of component masses is left for future work.
 
There could be ways in which our conclusions are on the pessimistic side. For example, a more physical parametrization of the EOS as in \cite{Lackey:2014fwa} allows one to fold in physical constraints such as causality, which is bound to improve parameter estimation. Moreover, it was recently found that the implementation of quantum squeezing in the interferometers may improve the measurability of tidal deformabilities by a few tens of percent \cite{PhysRevD.91.044032}. Finally, it is worth noting that with the ``plausible'' BNS detection rate of $\sim 40$ per year at design sensitivity \cite{Abadie:2010cf}, the desired number of sources could be collected over the course of a few years.

\section*{Acknowledgements}

MA, JM, MT, CVDB, and JV were supported by the research programme of the Foundation for Fundamental Research on Matter (FOM), which is partially supported by the Netherlands Organisation for Scientific Research (NWO). JV was also supported by the UK Science and Technology Facilities Council Grant No. ST/K005014/1. SV acknowledges the support of the National Science Foundation and the LIGO Laboratory. LIGO was constructed by the California Institute of Technology and Massachusetts Institute of Technology with funding from the National Science Foundation and operates under Cooperative Agreement No. PHY-0757058.  The work
was funded in part by a Leverhulme Trust research project grant. The authors would like to acknowledge the LIGO Data Grid clusters, without which the simulations could not have been performed. Specifically, these include the computing resources supported by National Science Foundation Grants No. PHY-0923409 and No. PHY-0600953 to the University of Winskonsin--Milwaukee. Also, we thank the Albert Einstein Institute in Hannover, Germany, supported by the Max-Planck-Gesellschaft, for use of the Atlas high-performance computing cluster. It is a pleasure to thank Brynmor Haskell for the useful input on the ``I-Love-Q'' relations and Benjamin D. Lackey for the useful discussions.

\section*{Appendix: Effect of the prior on component masses}
\label{appendix}

Unlike with our evidence calculations, in the case of posterior density functions it is relatively easy to ``reweight'' the sampling of parameter space so as to make $p(\vec{\theta}, \{c_j\} | d, I)$ correspond to different priors on the parameters \cite{Veitch:2014wba}. The degradation in the estimation of $c_0$ (and for that matter, hypothesis ranking) happened when we changed the way the component masses in the injections were distributed. Hence it is of interest to study the effect of the prior on the masses in particular. 

Let us pretend to have perfect knowledge of the astrophysical mass distribution -- in our example a Gaussian with $\mu_m  =1.35\, M_\odot$ and $\sigma_m = 0.05\, M_\odot$ -- and take the prior on $m_1$, $m_2$ to be identical to it. In the case of zero spins, the result is shown in Fig.~11. We see that the significant biases we encountered in Fig.~9 have largely gone away. In Fig.~12 we also include spins as before; here too, the biases seen earlier are strongly mitigated, though the larger parameter space to be probed still causes larger statistical errors. 

This is not a typical case of a ``prior-dominated'' inference on a parameter, since the bias originates from a bad choice of priors for different parameters ($m_1$, $m_2$) than the one that we are interested in ($c_0$).
Two important details that make this bad choice manifest itself as a bias in the $c_0$ posteriors are the following.  
First, there is the fact that the parameters $\lambda_A$, through which $c_0$ is inferred, have an implicit dependence on the masses $m_A$.
The $c_0$ posterior is determined by the posterior on the $m$--$\lambda$ plane for each component NS, and if the masses are biased then so is the inferred $\lambda(m)$ curve.
Second, since $c_0$ is treated as an independent parameter, the bias enters through the mass prior, in the process of marginalizing over $m_1$ and $m_2$, consistently for each source, and is therefore a persistent bias that will not average out as the number of sources increases.

In conclusion, the biases we see in the estimation of $c_0$ mostly result from the mismatch between the mass distribution for the sources and the prior distribution of component masses in the Bayesian analysis of the data. The relatively small remaining biases that occur when the injected mass distribution is the same as the prior can be attributed to the quadratic approximation for $\lambda(m)$ used in the template waveforms, and the fact that when most of the masses are in a narrow interval, less of the underlying tidal deformability function is being probed by the sources.

\begin{figure}
\label{fig:Gausspriornospin} 
\includegraphics[width=\columnwidth]{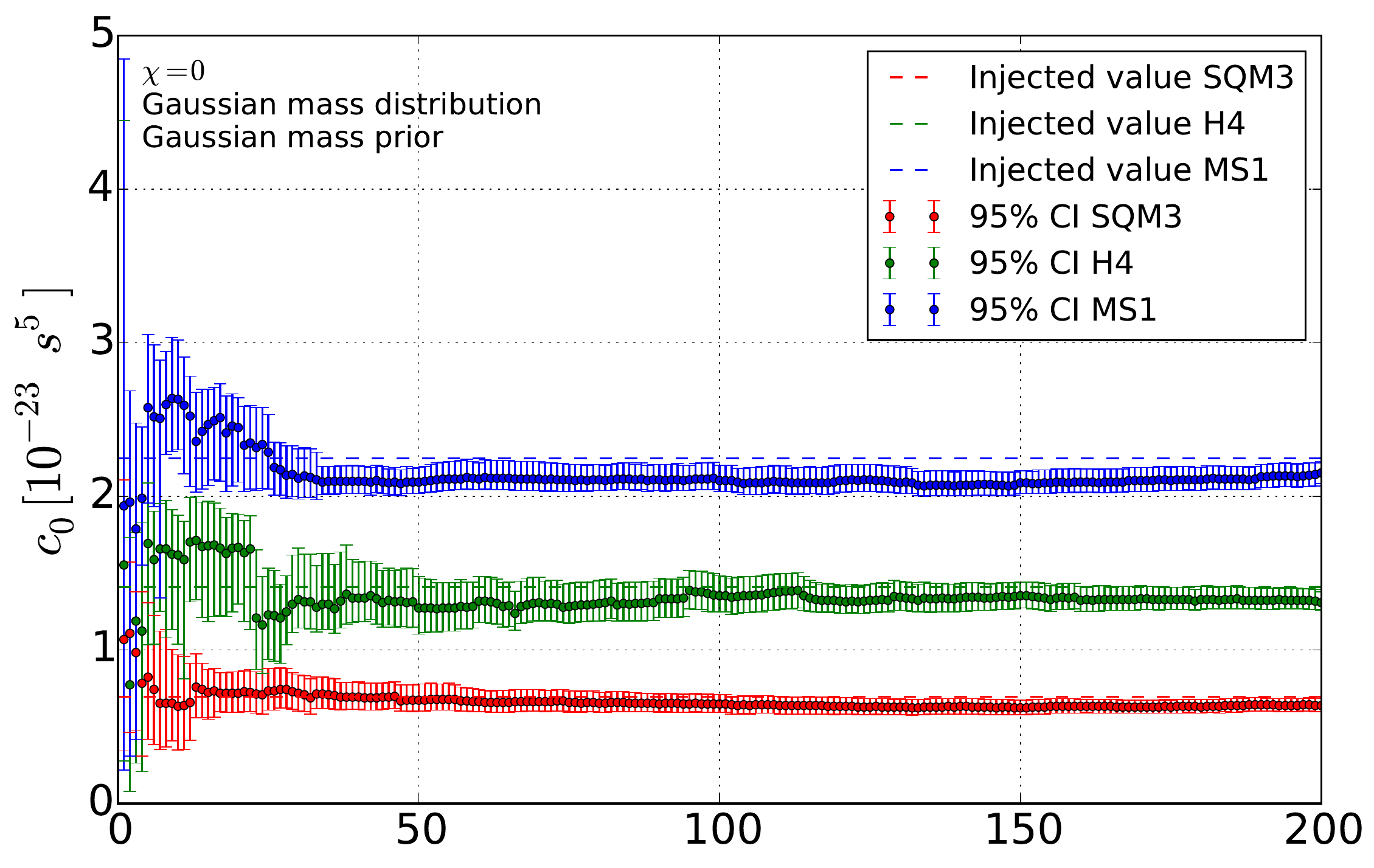} 
\\
\caption{The same as in Fig.~9, but this time using a Gaussian prior for the component masses that exactly matches the injected mass distribution. The significant biases that were seen before have largely disappeared.} 
\end{figure}

\begin{figure}
\label{fig:Gausspriorspin} 
\includegraphics[width=\columnwidth]{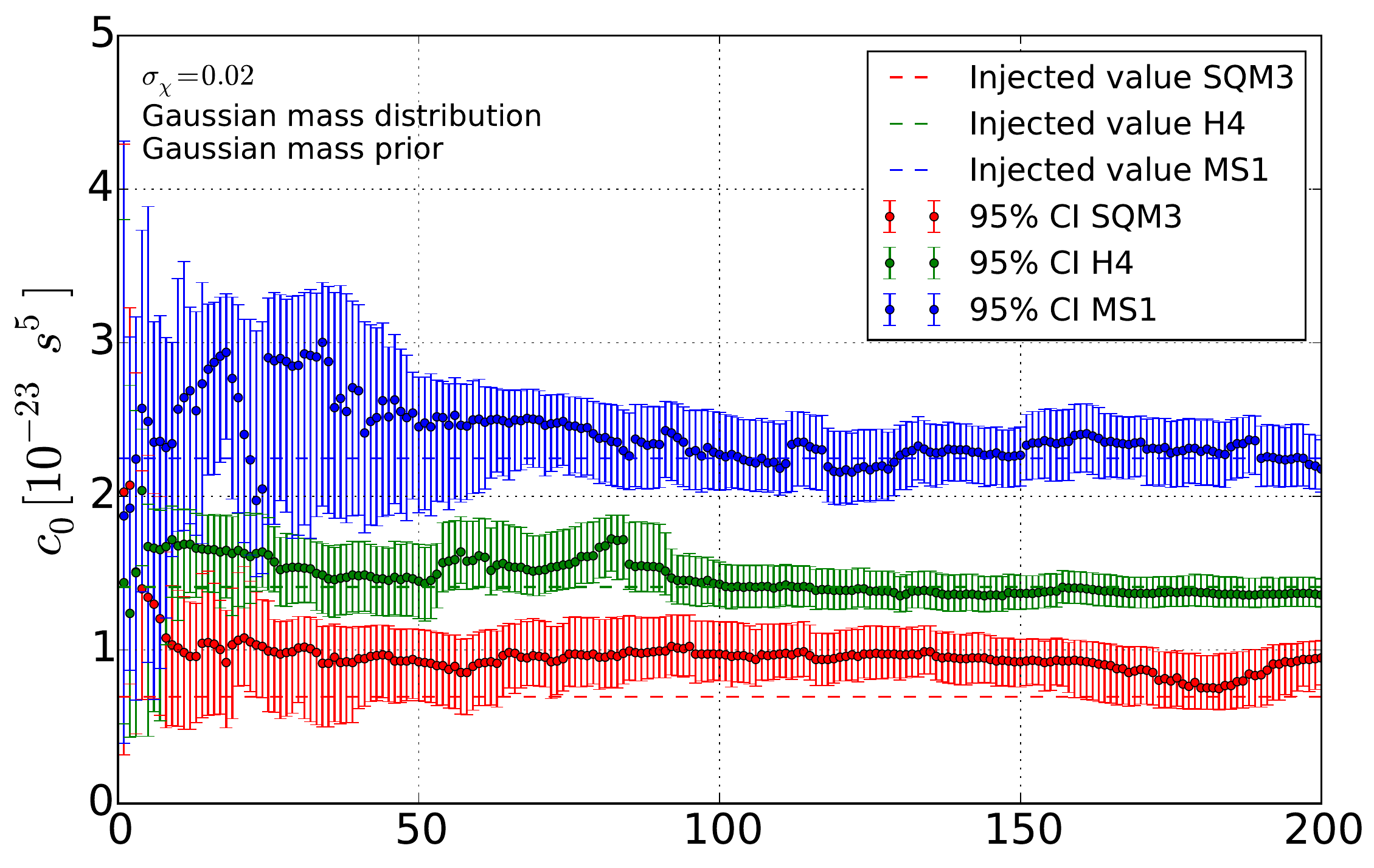} 
\\
\caption{The same as in Fig.~11, but now with spins switched on. Again we use a Gaussian prior for the component masses that matches the injected distribution. Here too, the biases have been mitigated.} 
\end{figure}

\vskip 10cm

\nocite{*}
\bibliography{BNS_EoS_v7}

\end{document}